\documentclass[11pt]{article}
\usepackage{graphicx,amssymb}
\newcommand{\SetFigFont}[3]{}

\title{Decay of Solutions of the Teukolsky Equation for Higher Spin in the Schwarzschild Geometry}
\author{F.\ Finster\thanks{Research supported in part by the Deutsche
Forschungsgemeinschaft.}, J.\ Smoller\thanks{Research supported in
part by the Humboldt Foundation and the National Science Foundation,
Grant No.~DMS-0603754.}}
\date{July 2007 / August 2016}

\newtheorem{Def}{Def.}[section]
\newtheorem{Thm}[Def]{Theorem}
\newtheorem{Prp}[Def]{Proposition}
\newtheorem{Lemma}[Def]{Lemma}

\newcommand{\Proof}{{\em{Proof. }}}
\newcommand{\QED}{\ \hfill $\FBox$ \\[1em]}
\newcommand{\spc}{\;\;\;\;\;\;\;\;\;\;}

\newcommand{\R}{\mathbb{R}}
\newcommand{\1}{\mbox{\rm 1 \hspace{-1.05 em} 1}}

\newcommand{\sR}{\mbox{\rm \scriptsize I \hspace{-.8 em} R}}
\newcommand{\N}{\mathbb{N}}

\newcommand{\beq}{\begin{equation}}
\newcommand{\eeq}{\end{equation}}

\newcommand{\FBox}{\rule{2mm}{2.25mm}}

\begin{document}
\maketitle

\begin{abstract}
We prove that the Schwarzschild black hole is linearly stable under
electromagnetic and gravitational perturbations. Our method is to show
that for spin~$s=1$ or~$s=2$, solutions of the Teukolsky equation with
smooth, compactly supported initial data outside the event horizon,
decay in~$L^\infty_{\mbox{\scriptsize{loc}}}$.
\end{abstract}

\tableofcontents

\section{Introduction}
\setcounter{equation}{0}
In polar coordinates~$(t,r,\vartheta, \varphi)$, the line element of the Schwarzschild
metric is given by
\[ ds^2 \;=\; g_{jk}\:dx^j x^k \;=\;
\frac{\Delta}{r^2} \: dt^2
\:-\: \frac{r^2}{\Delta}\:dr^2 - r^2\: d \vartheta^2 - r^2\: \sin^2 \vartheta\:
d\varphi^2\:, \]
where
\[ \Delta \;=\; r^2 - 2M r \:, \]
and~$M$ is the mass of the black hole. The zero~$r_1:=2M$ of~$\Delta$ defines
the event horizon of the black hole.
The evolution of a massless wave~$\Phi$ of general spin~$s$ in the Schwarzschild geometry is described by the Bardeen-Press equation~\cite{BP}.
Teukolsky~\cite{Teukolsky} later generalized the equation to the Kerr geometry, and therefore the equation is usually referred to as the {\em{Teukolsky equation}}~\cite{C}. We here work with a particularly convenient form of the
Teukolsky equation due to Whiting~\cite{W}:
\begin{eqnarray}
\lefteqn{ \left[ \partial_r \Delta \partial_r - \frac{1}{\Delta} \left( r^2 \partial_t - (r-M) s \right)^2
- 4 s r \partial_t \right. } \nonumber \\
&& \left. + \partial_{\cos \vartheta} \sin^2 \vartheta\:
\partial_{\cos \vartheta}
+ \frac{1}{\sin^2 \vartheta} \left(\partial_\varphi + i s \cos \vartheta \right)^2 \right] \Phi(t,r,\vartheta, \varphi)
\;=\; 0 \:. \label{teq}
\end{eqnarray}
This is a second order scalar wave equation, having {\em{complex}}
coefficients if~$s \neq 0$.
(Note that our wave function differs from the function~$\psi$ in~\cite{Teukolsky} by a power of~$\Delta$, $\Phi = \Delta^{\frac{s}{2}} \psi$.)
The parameter~$s$ is either integer or half-integer valued.
The case~$s=0$ gives the scalar wave equation.
Of particular physical interest are the cases~$s=\frac{1}{2}, 1, 2$,
which correspond respectively to the massless Dirac equation, Maxwell's equations and the equations for linearized gravitational waves.

In this paper, we consider the Cauchy problem for the Teukolsky equation~(\ref{teq}) with initial data
\beq \Phi_{|t=0}\;=\; \Phi_0\:, \qquad \partial_t \Phi_{|t=0} \;=\; \Phi_1 \:, \label{id} \eeq
which is smooth and compactly supported outside the event horizon.
Our main theorem is the first
rigorous result on time-dependent solutions of the Teukolsky equation for higher spin, and proves linear stability of the Schwarzschild black hole under
electromagnetic and gravitational perturbations.
\begin{Thm} \label{thm1} For spin~$s=1$ or~$s=2$, the solution of the Cauchy problem~(\ref{teq}, \ref{id})
for~$(\Phi_0, \Phi_1) \in C^\infty_0((r_1, \infty) \times S^2)^2$ decays
in~$L^\infty_{\mbox{\scriptsize{loc}}}((r_1, \infty) \times S^2)$ as~$t \rightarrow -\infty$.
\end{Thm}
The study of linear stability of the Schwarzschild geometry was initiated in 1957 by
Regge and Wheeler~\cite{RW}, who discussed mode stability for metric perturbations.
In the case~$s=0$, the Cauchy problem~(\ref{teq}, \ref{id}) was considered (for more general initial data) by
Kay and Wald \cite{KW}, and they obtained a time independent~$L^\infty$-bound.
Decay in~$L^\infty_{\mbox{\scriptsize{loc}}}$ was 
proved in~\cite{FKSY, FKSY2} in the Kerr geometry,
and worked out in~\cite{K} in the Schwarzschild geometry.
Related results for~$s=0$ in the Schwarzschild geometry
were obtained in~\cite{Price, DR}.
If~$s=\frac{1}{2}$, local decay was proved in the Kerr geometry in~\cite{FKSY03}
(for both the massive and massless case),
and an exact decay rate was given in the massive case~\cite{FKSYdecay}.
Up to now, for higher spin~$s=1$ (Maxwell's equations) and~$s=2$
(linearized gravitational waves) only mode analyses have been carried
out for the Teukolsky equation; see~\cite{PT} for a numerical study and~\cite{W} for a rigorous proof of mode stability. We also mention the
two papers on the Regge-Wheeler equation~\cite{FM, BS}.
In the first paper pointwise decay of solutions is proved, 
whereas in the second paper time integrals of solutions are estimated locally in space.

To consider the limit~$t \rightarrow -\infty$ (and not~$t \rightarrow
+\infty$) is purely a matter of convenience.
To see this, first note that in a general space-time, a massless field of spin~$s \neq 0$
satisfies a coupled system
of~$2s+1$ complex, first order partial differential equations.
As shown by Teukolsky~\cite{Teukolsky}, this system can be decoupled
in the Kerr geometry by multiplying with a suitable first order differential operator. Then the first
component of the system satisfies the Teukolsky equation~(\ref{teq}), whereas the
last component also satisfies~(\ref{teq}), but with~$s$ replaced by~$-s$.
From either the first or the last component, all the other components can be obtained by applying the so-called Teukolsky-Starobinsky identities, see~\cite{C}. In view of this, we may restrict attention to the Teukolsky equation~(\ref{teq})
for either~$s$ or~$-s$.
Next, we point out that
the Teukolsky equation~(\ref{teq}) is invariant under the
transformations~$(t,s, \vartheta,
\varphi) \rightarrow (-t, -s, \vartheta, -\varphi)$.
We thus see that Theorem~\ref{thm1} also makes a similar statement on the
solution~$\Phi$ of the Teukolsky equation for spin $-s$ in
the limit~$t \rightarrow +\infty$.
Since the Teukolsky-Starobinsky identitities relate
the solutions of the Teukolsky equations for~$\pm s$ to each other,
obtaining decay for the spin $-s$ equation as~$t \rightarrow \infty$
immediately yields decay for the spin $s$ equation as~$t \rightarrow \infty$.
Thus there is no loss in generality to consider in Theorem~\ref{thm1}
the case~$t \rightarrow -\infty$. This case will turn out to
be most convenient if one uses 
the sign conventions in~\cite{Teukolsky} as well
as the factor~$e^{-i \omega t}$ in the time separation.

Let us specify in which sense the Teukolsky equation governs
the linear perturbations of the Schwarzschild black hole.
For spin~$s=1$, the Teukolsky equation takes into account all electromagnetic
perturbations except for adding a constant electric charge
(see~\cite[p.\ 644]{Teukolsky}), thus perturbing Schwarzschild to the Reissner-Nordstr\"om
space-time.
For~$s=2$, the Teukolsky equation describes perturbations of the
Weyl tensor, and it is a quite difficult task to reconstruct
metric perturbations from a solution of the Teukolsky equation;
for details see~\cite{WL, WP}.
It is important to keep in mind that the Teukolsky equation
excludes perturbations of Schwarzschild to the Kerr space-time,
but does take into account all other regular perturbations (see~\cite{Wald}).
Hence our Theorem shows linear stability of the Schwarzschild black hole under all perturbations,
except for linear perturbations to the stationary Kerr-Newman black hole.

We now briefly discuss energy conservation and its role in our proof.
In the Schwarz\-schild geometry, the physical energy can be written
as the spatial integral of a {\em{positive energy density}}.
More precisely, in the cases~$s=0$ and~$1$, this energy is obtained by
integrating the normal component of the vector field~$T^i_{\;\;0}$,
where~$T_{ij}$ is the energy-momentum tensor corresponding to
the spin~$s$ field,
\[ E \;=\; \int_{t={\mbox{\scriptsize{const}}}}
T_{ij}\: \nu^i\: \left( \frac{\partial}{\partial t} \right)^j
d\mu \;=\; \int_{t={\mbox{\scriptsize{const}}}}
T^0_{\;\;\,0}\:d\mu \:, \]
where~$\nu$ is the future-directed normal and~$d\mu$ is the
integration measure on the hypersurface~$t={\mbox{const}}$.
This energy is conserved because the energy-momentum tensor
is divergence-free and~$\partial_t$ is a Killing field.
Using the dominant energy condition and the fact that~$\partial_t$ is
timelike, it is easy to verify that the energy density is
indeed non-negative.
In the case~$s=2$, a conserved energy of the gravitational field
is given by the integral of
the Bel-Robinson tensor~$Q$ (see for example~\cite[p.\ 42ff]{KN})
\beq \label{Edef}
E \;=\; \int_{t={\mbox{\scriptsize{const}}}} Q^0_{\;\;\,000} \:d\mu \:.
\eeq
This energy is the sum of the gravitational energy of the Schwarzschild
metric and the energy of the gravitational wave.
As it does not seem possible to isolate the energy of the gravitational wave
from this expression, it seems preferable to proceed differently as follows:

Given a solution~$\Phi$ of the Teukolsky equation for~$s=2$, all the components of the Weyl
tensor can be computed using the Teukolsky-Starobinsky identities (see~\cite{C}).
The corresponding metric perturbations can be obtained by integration
(for explicit formulas see~\cite{WL}).
Expanding the resulting metric perturbations in spherical harmonics and choosing
a specific gauge (as worked out in detail in~\cite{RW, zerilli}), one obtains
solutions of the well-known Regge-Wheeler and Zerilli equations.
For each angular mode, the corresponding energy of the gravitational wave
is given as the spatial integral over a definite energy density
(this energy density is given in a convenient form in~\cite[eqn~(1.10)]{FM}).
Taking the sum of all modes, one obtains a corresponding formula for the energy
of the gravitational wave described by~$\Phi$.
Clearly, the resulting expressions for the energy density are
very complicated and involve higher derivatives of~$\Phi$.
For this reason, we are unable to
use the explicit form of the energy density.
In particular, we cannot work with a corresponding energy scalar
product. As a consequence, the associated Hamiltonian will
not be a symmetric operator, and thus we cannot use spectral theory
in Hilbert spaces. The main technical difficulty
of the present paper is to prove
completeness and decay without using the spectral theorem.
Nevertheless, we will make use of the existence of a positive energy density
a few times, without refering to its explicit form.

The main step in the proof is to derive an integral representation for the
propagator, whereby we integrate over the real line~$\R$ together with
another line parallel to~$\R$ (see Theorem~\ref{thmintrep2}).
The latter line integral reflects the fact that the
essential spectrum of the Hamiltonian has a contribution in the complex plane.
The appearance of a complex essential spectrum can be understood from the fact that the Teukolsky equation~(\ref{teq}) involves first order time
derivatives, which after time separation~$e^{-i \omega t}$ with real~$\omega$
lead to complex potentials in the resulting radial equation. 
These complex potentials make it impossible to apply
standard techniques used for 1-dimensional Schr\"odinger equations.
In particular, the fundamental solutions of the radial ODE behave
asymptotically near infinity like a power of~$r$ times a plane wave
(and no longer just like plane waves as in the case~$s=0$).
This requires us to develop new techniques like obtaining
refined WKB estimates for Jost solutions with complex potentials,
and working with non-closed integration contours.
To prove completeness, we use an idea in Bachelot~\cite{B},
which reduces the completeness problem to obtaining resolvent estimates
for large values of the spectral parameter~$\omega$.

We remark that in the case~$s=\frac{1}{2}$, Theorem~\ref{thmintrep2}
gives an integral representation for the propagator of the massless Dirac equation,
which is considerably different from the integral representation obtained in~\cite{FKSY03}.
This surprising fact will be discussed in Section~\ref{sec6}.

This paper is organized as follows. In Section~\ref{sec2} we separate out the
time and angular dependence in the Teukolsky equation and obtain the radial ODE.
In Section~\ref{sec3} we construct holomorphic families of Jost solutions of the radial equation
which have prescribed asymptotics near the event horizon or at infinity.
In Section~\ref{sec4} we write the Teukolsky equation in Hamiltonian form
and express the resolvent of the Hamiltonian in terms of the Jost functions.
Section~\ref{sec4a} is devoted to the derivation of WKB estimates,
which give precise bounds for the Jost functions asymptotically near the event
horizon and at infinity, and also globally if~$|\omega|$ is sufficiently large.
Using these estimates, in Section~\ref{secres} we study the decay properties
of the resolvent for large values of the spectral parameter.
These resolvent estimates allow us in Section~\ref{seccom}
to express the propagator in terms of
contour integrals, thereby also obtaining a completeness result.
In Section~\ref{seccontour} we use classical Whittaker functions together with
an energy argument to show that the integration contour can be deformed onto
both the real axis and another line parallel to it.
In Section~\ref{secdecay} we prove Theorem~\ref{thm1} using
a Riemann-Lebesgue argument for a finite number of angular modes,
together with an estimate for the remaining infinite number of modes.
Finally, Section~\ref{sec6} is devoted to general remarks on our integral representation
in the case~$s=\frac{1}{2}$ and on possible extensions of our methods
to the Kerr geometry.

\section{Separation of Variables} \label{sec2}
\setcounter{equation}{0}
Using spherical symmetry, we can separate out the angular dependence with
the usual multiplicative ansatz
\[ \Phi(t,r,\vartheta, \varphi) \;=\; R(t,r)\: Y(\vartheta, \varphi)\:. \]
The spin-weighted spherical harmonics~$Y=\:_sY_{lm}$ with $l=|s|,|s|+1,\ldots$,
$m=-l,-l+1,\ldots, l$ (see~\cite{spinweight}) form an eigenvector basis of the
angular operator
\[ {\mathcal{A}} \;=\; -\partial_{\cos \vartheta} \sin^2 \partial_{\cos \vartheta}
- \frac{1}{\sin^2 \vartheta} \left(\partial_\varphi + i s \cos \vartheta \right)^2 \:, \]
on~$L^2(S^2)$, corresponding to the eigenvalues~$\lambda_l=l(l+1)-s^2$
(note that our angular operator is related to the operator~$\mathfrak{H}_0$ in~\cite{PT}
by~${\mathcal{A}} = -{\mathfrak{H}}_0 - s^2$).
Restricting attention to one angular momentum mode, the Teukolsky equation reduces to
\beq \label{teukolskysep}
\left[ \partial_r \Delta \partial_r - \frac{1}{\Delta} \left( r^2 \partial_t - (r-M) s \right)^2
- 4 s r \partial_t \:-\: \lambda \right] R(t,r) \;=\; 0\:,
\eeq
where we set $\lambda=\lambda_l$. We transform to the Regge-Wheeler variable~$u \in \R$ defined by
\beq \label{RW}
\frac{du}{dr} \;=\; \frac{r^2}{\Delta} \:,\qquad {\mbox{so}} \qquad
u \;=\; r + 2M\, \log(r-2M) \:,
\eeq
which maps the event horizon~$r=2M$ to~$u=-\infty$. Furthermore,
setting
\[  \phi(t,r) \;=\; r\, R(t,r) \:, \]
the Teukolsky equation becomes
\[ \left[ \frac{r^3}{\Delta} \partial_u r^2 \partial_u \frac{1}{r} - \frac{1}{\Delta} \left( r^2 \partial_t - (r-M) s \right)^2
- 4 s r \partial_t \:-\: \lambda \right] \phi(t,r) \;=\; 0\:. \]
Applying the identity~$\partial_u r^2 \partial_u = r \partial_u^2 r - r (\partial_u^2 r)$, we can
write this equation in the simpler form
\beq \label{teq2}
\left[ \partial_u^2 -\left( \partial_t - \frac{(r-M) s}{r^2} \right)^2
- \frac{4 s \Delta}{r^3}\,  \partial_t - \frac{\partial_u^2 r}{r}
- \lambda\: \frac{\Delta}{r^4} \right] \phi(t,r) \;=\; 0\:.
\eeq

Using the time translation symmetry, we can further separate out the time dependence
with the ansatz
\beq \label{tsep}
\phi(t,r) \;=\; e^{-i \omega t}\: \phi(r)\:.
\eeq
Then the Teukolsky equation reduces to the ODE
in Schr\"odinger form
\beq \label{schroedinger}
-\frac{d^2}{du^2} \,\phi(u) + V(u)\, \phi(u) \;=\; 0 \:,
\eeq
where the potential~$V$ is given by
\beq \label{Vdef}
V(u) \;=\; -\omega^2 + is \omega \left[ \frac{2 (r-M)}{r^2} - \frac{4 \Delta}{r^3}
\right] + \frac{(r-M)^2\, s^2}{r^4} + \frac{\partial_u^2 r}{r} + \lambda\: \frac{\Delta}{r^4}\:.
\eeq
Note that in the case~$s \neq 0$, $V$ is {\em{complex}} even for real~$\omega$.

\section{Construction of the Jost Solutions} \label{sec3}
\setcounter{equation}{0}
In this section we construct Jost solutions~$\acute{\phi}$ and~$\grave{\phi}$
of the Schr\"odinger equation~(\ref{schroedinger}), which are defined
by their asymptotic behavior near the event horizon and near infinity,
respectively. Near the event horizon, the potential has the limit
\[ \lim_{u \rightarrow -\infty} V(u) \;=\; - \Omega^2
\qquad {\mbox{where}} \qquad
\Omega \;:=\; \omega - \frac{i s}{4M}\:. \]
Writing the Schr\"odinger equation in the form
\[ (-\partial_u^2 - \Omega^2)\, \acute{\phi} \;=\; -W(u)\, \acute{\phi}(u) \:, \]
the potential~$W$ behaves near the event horizon linearly in $(r-r_1)$, and thus has exponential decay in the Regge-Wheeler coordinate~(\ref{RW})
near~$u=-\infty$. More precisely, for~$u$ near~$-\infty$
there is a constant~$c>0$ such that
\[ |W(u)| \;\leq\; c\: e^{\frac{u}{2M}}\:. \]
Using this exponential decay, $\acute{\phi}$ can be constructed
 exactly
as in~\cite[Theorem~3.1]{FKSY2} (see also~\cite{AR, K}).
The properties of~$\acute{\phi}$ are summarized in the following theorem.
\begin{Thm} \label{thm31}
For every~$\omega$ in the domain
\[ D_- \;=\; \left\{ \omega \:\Big|\:
 {\mbox{\rm{Im}}}\, \omega \;<\; \frac{s}{4M} + \frac{1}{4M} \right\} \]
there is a solution~$\acute{\phi}_-$ of~(\ref{schroedinger})
having the asymptotics
\beq \label{acutephiasy}
\lim_{u \to -\infty} e^{-i \Omega u} \:\acute{\phi}_-(u) \;=\; 1 \:,\spc
\lim_{u \to -\infty} \left(e^{-i \Omega u} \:\acute{\phi}_-(u) \right)' \;=\; 0 \:.
\eeq
These solutions can be chosen to form a holomorphic family, in the sense that for every~$u \in \R$,
the function~$\acute{\phi}_-(u)$ is holomorphic in~$\omega \in D_-$.
Similarly, on the domain
\[ D_+ \;=\; \left\{ \omega \:\Big|\:
 {\mbox{\rm{Im}}}\, \omega \;>\; \frac{s}{4M} - \frac{1}{4M} \right\} \]
there is a holomorphic family of solutions~$\acute{\phi}_+$ of~(\ref{schroedinger})
with the asymptotics
\[ \lim_{u \to -\infty} e^{i \Omega u} \:\acute{\phi}_+(u) \;=\; 1 \:,\spc
\lim_{u \to -\infty} \left(e^{i \Omega u} \:\acute{\phi}_+(u) \right)' \;=\; 0 \:. \]
\end{Thm}

\vspace*{1em}
Near infinity, the potential has the following asymptotic form,
\beq \label{Vasy}
V(u) \;=\; -\omega^2 - \frac{2is \omega}{u} \:+\: {\mathcal{O}}\left( \frac{\log u}{u^2} \right) .
\eeq
In the remainder of this section we always assume that~$u \gg 1$.
In the case~$s=0$, the solutions~$\grave{\phi}$ were constructed
in~\cite{FKSY2, K}; thus we assume in what follows that~$s \geq \frac{1}{2}$.
Because of the non-integrable $u^{-1}$-term in~$V$, the standard Jost solution
method~\cite{AR} cannot be implemented.
We choose~$u_0$ so large that~$V$ has no zeros on~$[u_0, \infty)$. We
introduce the WKB wave functions~$\acute{\alpha}$ and~$\grave{\alpha}$ by
\beq \label{adef}
\acute{\alpha}(u) \;=\; \acute{c}\,V(u)^{-\frac{1}{4}} \exp \left(\int_{u_0}^u \sqrt{V} \right) ,\qquad
\grave{\alpha}(u) \;=\; \grave{c}\, V(u)^{-\frac{1}{4}} \exp \left(-\int_{u_0}^u \sqrt{V} \right) ,
\eeq
with constants~$\acute{c}, \grave{c} \neq 0$ to be determined later.
To explain the sign convention for~$\sqrt{V}$, we first note that
taking the square root of~(\ref{Vasy}) gives
\beq \label{sqVasy}
\sqrt{V(u)} \;=\; \pm \left( i \omega - \frac{s}{u} \right)
\:+\: {\mathcal{O}}\left( \frac{\log u}{u^2} \right) \:.
\eeq
Our sign convention is
\beq \label{signconvention}
\left\{ \begin{array}{cl} + & {\mbox{if~${\mbox{Im}}\, \omega \leq 0$}} \\
- & {\mbox{if~${\mbox{Im}}\, \omega > 0$}} . \end{array} \right.
\eeq
Thus if~${\mbox{Im}}\, \omega \neq 0$, 
the real part of $\sqrt{V(u)}$ is positive for large~$u$ and so~$\grave{\alpha}$ decays at plus infinity.
Furthermore, we note that our sign convention does not change if~$\omega$ approaches the real line from below.
Also, we point out that for real~$\omega$, the function~$\grave{\alpha}$ does not decay at infinity,
but increases polynomially like~$u^s$.

The functions~$\acute{\alpha}$ and~$\grave{\alpha}$ are solutions of the equation
\beq \label{Leq}
L \alpha \;=\; 0 \:,
\eeq
where~$L$ is the differential operator defined by
\[ L \;=\; -\partial_u^2 + V_0
\qquad {\mbox{and}} \qquad
V_0 \;:=\; V -\frac{V''}{4 V} + \frac{5}{16}
\left( \frac{V'}{V} \right)^2 \:. \]
Writing the Schr\"odinger equation~(\ref{schroedinger}) as
\beq \label{sch2}
L\, \phi \;=\; -W\, \phi \:,
\eeq
we see that the new potential~$W := V-V_0$ is integrable, since
\beq \label{Wbound}
|W(u)| \;\leq\; \frac{c}{1+|\omega|} \: \frac{1}{u^3}\qquad {\mbox{on $[u_0, \infty)$}}\:.
\eeq

Since the WKB wave functions~$\acute{\alpha}$ and~$\grave{\alpha}$ form a
fundamental system for~(\ref{Leq}), we can use them to construct a
Green's function for the operator~$L$. In what follows~$\Theta$ denotes the
usual Heaviside function.
\begin{Lemma} Under the sign convention~(\ref{signconvention}), the function
\beq \label{grdef}
S(u,v) \;=\; \frac{1}{2} \:\Theta(v-u)\, (V(u) V(v))^{-\frac{1}{4}}
\left[ \exp \left( \int_u^v \sqrt{V} \right) - \exp \left(- \int_u^v \sqrt{V} \right) \right]
\eeq
is, for all~$u,v > u_0$, a distributional solution of the equation
\beq \label{green}
L_u\, S(u,v) \;=\; \delta(u-v)\:.
\eeq
\end{Lemma}
{\Proof} We make the ansatz
\[ S(u,v) \;=\; \Theta(v-u) \left( c_1(v)\: \acute{\alpha}(u)
\:+\: c_2(v)\: \grave{\alpha}(u) \right) \]
and determine the coefficients~$c_1$ and~$c_2$ from the conditions
\[ \lim_{u \nearrow v} S(u,v) \;=\; 0 \:,\qquad
\lim_{u \nearrow v} \partial_u S(u,v) \;=\; -1\:. \]
This gives~(\ref{grdef}), and a straightforward calculation yields~(\ref{green}).
\QED

We now make the perturbation ansatz
\beq \label{pert}
\grave{\phi} \;=\; \sum_{n=0}^\infty \phi^{(n)}\:,
\eeq
where the~$\phi^{(n)}$ are defined by the iteration scheme
\beq \label{pert2}
\left. \begin{array}{rcl}
\phi^{(0)}(u) &=& \grave{\alpha}(u) \\
\phi^{(l+1)}(u) &=& \displaystyle -\int_u^\infty S(u,v)\:
W(v)\: \phi^{(l)}(v)\: dv \:. \end{array} \right\}
\eeq
Before stating the next theorem, we must study
the asymptotics of~$\grave{\alpha}$ near infinity.
Carrying out the integral in~(\ref{adef}) using~(\ref{sqVasy}),
we obtain
\[ \grave{\alpha}(u) \;\sim\; e^{\mp i \omega u \pm s \log u }
\;=\; u^{\pm s}\: e^{\mp i \omega u}\:. \]
Due to our sign convention~(\ref{signconvention}), we find that
\beq \label{aasy}
\grave{\alpha}(u) \;\sim\;
\left\{ \begin{array}{cl} u^{-s}\: e^{i \omega u} & {\mbox{if
${\mbox{Im}}\, \omega \geq 0$}} \\
u^s\: e^{-i \omega u} & {\mbox{if
${\mbox{Im}}\, \omega < 0$}}. \end{array} \right.
\eeq
We next prove that the perturbation series~(\ref{pert})
converges to a solution~$\grave{\phi}$ of the full
equation~(\ref{sch2}) having the same asymptotics as~$\grave{\alpha}$.

\begin{Thm} \label{thm34}
On the domain $E_+ := \{ \omega \:|\: \omega \neq 0 {\mbox{ and }} {\mbox{\rm{Im}}}\, \omega > 0 \}$,
there is a family of solutions~$\grave{\phi}_+(u)$ of~(\ref{sch2}),
holomorphic in the interior of~$E_+$, having the asymptotics
\beq \label{phiasy0}
\lim_{u \rightarrow \infty} u^s\, e^{-i \omega u}\: \grave{\phi}_+(u) \;=\; 1\:,\spc
\lim_{u \rightarrow \infty} \left( u^s\, e^{-i \omega u}\: \grave{\phi}_+(u)\right)' \;=\; 0 \:.
\eeq
Likewise, on the domain $E_- := \{ \omega \:|\: \omega \neq 0 {\mbox{ and }} {\mbox{\rm{Im}}}\, \omega < 0 \}$,
there is a family of solutions~$\grave{\phi}_-(u)$ of~(\ref{sch2}), holomorphic
in the interior of~$E_-$, with the asymptotics
\beq \label{phiasy}
\lim_{u \rightarrow \infty} u^{-s}\, e^{i \omega u}\: \grave{\phi}_-(u) \;=\; 1\:,\spc
\lim_{u \rightarrow \infty} \left( u^{-s}\, e^{i \omega u}\: \grave{\phi}_-(u)\right)' \;=\; 0 \:.
\eeq
\end{Thm}
{\Proof} From the definitions~(\ref{adef}, \ref{grdef}) and
our sign convention~(\ref{signconvention}), it is obvious that
\begin{eqnarray*}
|\grave{\alpha}(u)| &\leq& d \:|V(u)|^{-\frac{1}{4}}\:
\exp \left( -\int_{u_0}^u |{\mbox{Re}}\, \sqrt{V}| \right) \\
|S(u,v)| &\leq& \Theta(v-u)\: |V(u)\, V(v)|^{-\frac{1}{4}}\:
\exp \left(\int_{u}^v |{\mbox{Re}}\, \sqrt{V}| \right)
\end{eqnarray*}
where we take
\[ d \;=\; \exp \left( 2 \int_{u_0}^{u_1} |{\mbox{Re}}\, \sqrt{V}| \right) , \]
with~$u_1$ chosen so large that the real part of the square root of V is positive for all $u>u_1$.

We will show inductively that for~$u>u_0$,
\beq \label{induct}
\left| \phi^{(l)}(u)\right| \;\leq\; 
\grave{c} d\:\frac{C^l}{u^{2l}\: l!} \:|V(u)|^{-\frac{1}{4}}\:
\exp \left( -\int_{u_0}^u |{\mbox{Re}}\, \sqrt{V}| \right) \:,
\eeq
where
\[ C \;=\; \frac{c}{2 (1+|\omega|)} \]
and~$c$ is as in~(\ref{Wbound}).
The case~$l=0$ is obvious. Assume that~(\ref{induct}) holds for a given~$l$.
Then
\begin{eqnarray*}
|\phi^{(l+1)}(u)| &\leq& \grave{c} d\:\frac{C^l}{l!}
|V(u)|^{-\frac{1}{4}}\:e^{-\int_{u_0}^u |{\mbox{\scriptsize{Re}}}\, \sqrt{V}|}
\int_u^\infty\frac{c}{1+|\omega|}\: \frac{1}{v^{3+2l}}\:dv \\
&=& \grave{c} d\:\frac{C^{l+1}}{u^{2(l+1)}\, (l+1)!}
|V(u)|^{-\frac{1}{4}}\:e^{-\int_{u_0}^u |{\mbox{\scriptsize{Re}}}\, \sqrt{V}|}\:.
\end{eqnarray*}

The estimate~(\ref{induct}) shows that the series~(\ref{pert})
converges absolutely, uniformly for $u>u_0$.
Similarly, one can show that the series obtained by diffentiating~(\ref{pert})
termwise again converges in the same sense.
Thus we can differentiate the series termwise, thereby showing
that~$\grave{\phi}$ is a solution of~(\ref{sch2}).
According to~(\ref{aasy}), we can choose~$\grave{c}$ such
that the function~$\phi^{(0)}$
satisfies the boundary conditions~(\ref{phiasy0}) or~(\ref{phiasy}),
and since the estimate~(\ref{induct}) involves a
factor~$u^{-2l}$, it is obvious that~$\grave{\phi}$
also satisfies the first relation in~(\ref{phiasy0}) or~(\ref{phiasy}).
The second relations are obtained by differentiating~(\ref{pert})
and~(\ref{pert2}) with respect to~$u$; a lengthy but straightforward
calculation yields the result.

To prove analyticity in~$\omega$, we first note that
$\grave{\alpha}$, $W$ and $S(u,v)$ are obviously analytic.
Hence each~$\phi^{(l)}$ is analytic being an integral
of analytic functions. Since the constants~$\grave{c}$ and~$d, C$
in~(\ref{induct}) can be chosen locally uniformly in~$\omega$,
we conclude from Morera's theorem that~$\grave{\phi}$
is also analytic.
\QED

\section{Hamiltonian Formulation, Construction of the Resolvent} \label{sec4}
\setcounter{equation}{0}
At this stage, we do not know whether the separation
ansatz~(\ref{tsep}) will give us a complete set of solutions of
the time-dependent equation~(\ref{teq2}). To remedy this situation, we
shall write the equation in Hamiltonian form. To this end we set
\[ \Psi \;=\; \left( \!\!\begin{array}{c} \phi(t,r) \\  i \partial_t \phi(t,r) \end{array} \!\!\right) \]
and obtain
\beq \label{Hdef}
i \partial_t \Psi \;=\; H  \Psi \qquad {\mbox{with}} \qquad H \;=\; \left( \begin{array}{cc} 0 & 1 \\ \alpha & \beta \end{array} \right)
\eeq
and
\begin{eqnarray}
\alpha &=& -\partial_u^2 + \frac{(r-M)^2\, s^2}{r^4} + \frac{\partial_u^2 r}{r} + \lambda\: \frac{\Delta}{r^4}
\label{alphadef} \\
\beta &=& i s \left[ \frac{2 (r-M)}{r^2} - \frac{4 \Delta}{r^3}
\right] . \label{betadef}
\end{eqnarray}
The Hamiltonian~$H$ can be considered as an operator on the Hilbert
space
\[ {\mathcal{H}} \;:=\; H^{1,2}(\R, du) \oplus L^2(\R, du)\:, \]
densely defined on the domain~${\mathcal{D}}(H) = {\mathcal{S}}(\R)^2$,
the Schwartz functions.
Note that~$H$ is {\em{not}} symmetric on~${\mathcal{H}}$.

We assume for the rest of this section that
\beq \label{oas}
{\mbox{Im}}\, \omega \not \in \left[ 0, \frac{s}{4M} \right] \:.
\eeq
For a given~$\omega$ satisfying these conditions,
we will show that the resolvent of~$H$ exists, and we will
express it in terms of the Jost solutions.
Depending on the sign of~${\mbox{Im}}\, \omega$, we let~$\grave{\phi}$
be the function~$\grave{\phi}_+$ or~$\grave{\phi}_-$ of Theorem~\ref{thm34}, respectively.
If~$\omega \in D_+ \cap D_-$, there are two Jost solutions~$\acute{\phi}_\pm$ near
the event horizon, one of which decays exponentially, the other of which grows exponentially.
We choose the solution with exponential decay. Thus in the case~${\mbox{Im}}\, \omega > s/(2M)$
we let~$\acute{\phi}=\acute{\phi}_+$, whereas in the case~${\mbox{Im}}\, \omega < s/(2M)$
we let~$\acute{\phi}=\acute{\phi}_-$.

The next lemma relies crucially on Whiting's mode stability result~\cite{W}.
\begin{Lemma} \label{lemmawcomplex} For any~$\omega$ satisfying~(\ref{oas}), the Wronskian
\beq \label{wronski}
w(\acute{\phi}, \grave{\phi}) \;:=\; \acute{\phi}' \,\grave{\phi} - 
\acute{\phi} \,\grave{\phi}'
\eeq
is non-zero.
\end{Lemma}
{\Proof}
If the Wronskian were zero, the solutions~$\acute{\phi}$ and~$\grave{\phi}$ would be linearly dependent.
Then there would be a solution~$\phi$ decaying exponentially fast
at both $u=\pm \infty$. If~${\mbox{Im}}\, \omega<0$, such solutions
have been ruled out by Whiting~\cite{W}. (Note that Whiting
considers the case~$s<0$ in the limit~$t \rightarrow +\infty$.
Using the symmetries~$(t,s, \vartheta, \varphi) \rightarrow
(-t, -s, \vartheta, -\varphi)$, this is equivalent to considering
the case~$s>0$ in the limit~$t \rightarrow -\infty$, which corresponds
to mode solutions in the lower half plane as considered here).

In the case~${\mbox{Im}}\, \omega > s/(4M)$ and~${\mbox{Re}}\, 
\omega \neq 0$, we use
that~$\phi$ is a solution of~(\ref{schroedinger}) to obtain
\[ 0 \;=\; \left\langle \left(-\frac{d^2}{du^2} + V \right) \phi, \phi
 \right\rangle_{L^2(\R)} -
\left\langle \phi, \left(-\frac{d^2}{du^2} + V \right) \phi
 \right\rangle_{L^2(\R)} . \]
Using the exponential decay of~$\phi$ as~$u \rightarrow \pm \infty$,
we can integrate by parts to get
\beq \label{contra}
0 \;=\; -2\: \left\langle \phi, ({\mbox{Im}}\, V)\, \phi
 \right\rangle_{L^2(\R)} .
\eeq
On the other hand, we see from~(\ref{Vdef}) that
\beq \label{ImV}
{\mbox{Im}}\, V \;=\; -2\, {\mbox{Re}}\, \omega
\left( {\mbox{Im}}\, \omega \:-\: \frac{s}{2} \left[ \frac{2 (r-M)}{r^2} - \frac{4 \Delta}{r^3} \right] \right) .
\eeq
A short computation shows that the round bracket is strictly positive.
Thus~${\mbox{Im}}\, V$ is either strictly positive or strictly negative,
contradicting~(\ref{contra}).

In the final case~${\mbox{Im}}\, \omega > s/(4M)$ and~${\mbox{Re}}\, 
\omega = 0$, we see from~(\ref{ImV}) that~$V$ is real,
and a short computation using~(\ref{Vdef}) shows that it is
even strictly positive. Using that according to~(\ref{acutephiasy}),
the fundamental solution~$\acute{\phi}$
is positive and increasing near~$u=-\infty$, we conclude that~$\acute{\phi}$ is convex. Hence it cannot be a multiple of the function~$\grave{\phi}$, which decays at infinity according to~(\ref{phiasy0}).
\QED
This lemma allows us to introduce the Green's function~$G(u,v)$ of the Schr\"odinger
equation~(\ref{schroedinger}) by the standard formula
\beq \label{Gdef}
G(u,v) \;=\; \frac{1}{w(\acute{\phi}, \grave{\phi})} \:\times\:
\left\{ \begin{array}{cl} \acute{\phi}(u)\, \grave{\phi}(v) & {\mbox{if~$v \geq u$}} \\
\grave{\phi}(u)\, \acute{\phi}(v) & {\mbox{if~$v < u$}}. \end{array} \right. .
\eeq
It satisfies the distributional equation
\beq \label{Geq}
\left( -\frac{d^2}{du^2} + V(u) \right) G(u,v) \;=\; \delta(u,v)\:.
\eeq
We let~$G$ denote the corresponding operator with integral kernel~$G(u,v)$.

\begin{Lemma} \label{lemma42}
For every~$\omega$ satisfying~(\ref{oas}),
$G$ is a bounded linear operator from~$L^2(\R)$ to~$H^{1,2}(\R)$, and
maps~$C^\infty_0(\R)$ to~${\mathcal{S}}(\R)$.
\end{Lemma}
{\Proof}
We restrict attention to the case~${\mbox{Im}}\, \omega < 0$,
since the other case is analogous.
To prove the first part, we let~$\psi$ be in~$L^2(\R)$.
Then the function~$G \psi$ can be written as
\[ (G \psi)(u) \;=\; \frac{1}{w(\acute{\phi}, \grave{\phi})} \left(
\acute{\phi}(u) \int_u^\infty \grave{\phi}(v)\, \psi(v)\, dv \:+\:
\grave{\phi}(u) \int_{-\infty}^u \acute{\phi}(v)\, \psi(v)\, dv \right) . \]
We consider only the first term, because the second term can be
treated similarly. Thus our task is to bound the function
\beq \label{fdef}
f(u) \;:=\; \acute{\phi}(u) \int_u^\infty \grave{\phi}(v)\, \psi(v)\, dv
\eeq
in~$H^{1,2}$. From Theorem~\ref{thm31} we know that
the solution~$\acute{\phi}$ behaves near the event horizon
like $\acute{\phi} \sim e^{i \Omega u}$. Integrating the Wronskian
equation~$\acute{\phi}' h - \acute{\phi} h'=1$ via the method
of variation of constants, we obtain another fundamental solution
\[ h(u) \;=\; \acute{\phi}(u) \int_u^0 \frac{1}{\acute{\phi}^2(x)}\:dx \:. \]
Using the asymptotics of~$\acute{\phi}$, one sees that~$h$
is bounded near the event horizon by a multiple of~$|e^{-i \Omega u}|$.
We conclude that the function~$\grave{\phi}$, being a linear combination of these two fundamental
solutions, satisfies the inequality
\[ |\grave{\phi}(v)| \;\leq\; C\, e^{-|{\mbox{\scriptsize{Im}}}\, \Omega|\, v} \qquad
{\mbox{for $v \ll 0$}}\:. \]
Using similar arguments near infinity, we conclude from Theorem~\ref{thm34} that
\[ |\acute{\phi}(u)| \;\leq\; C\, u^s \,e^{|{\mbox{\scriptsize{Im}}}\, \omega|\, u}
\qquad {\mbox{for $u \gg 0$}}\:. \]
Combining these inequalities with Theorems~\ref{thm31} and~\ref{thm34}, we have
estimates for both~$\acute{\phi}$ and~$\grave{\phi}$ at both asymptotic ends.
Since on any compact set, the solutions can be bounded using simple Gronwall
estimates, one sees
that choosing $\varepsilon = \min(|{\mbox{Im}}\, \omega|, |{\mbox{Im}}\, \Omega|)$,
we have the following estimate for sufficiently large~$c$,
\beq \label{wes}
\left| \acute{\phi}(u) \,\grave{\phi}(v) \right| \;\leq\;  c\: e^{-\varepsilon\, (v-u)}
\spc {\mbox{for all~$v \geq u$}}\:.
\eeq
This estimate gives the pointwise bound
\beq \label{fbound}
|f(u)| \;\leq\; c \int_u^\infty e^{-\varepsilon (v-u)}\: |\psi(v)|\; dv\:.
\eeq
Setting
\[ g(x) \;=\; c\,\Theta(x)\: e^{-\varepsilon x}\:, \]
we can write the right of side of~(\ref{fbound}) as
the convolution~$g * |\psi|$. Then using the Plancherel Theorem
together with the fact that convolution in position space corresponds
to multiplication in momentum space, we have
\[ \|f\|_2 \;\leq\; \|g * |\psi|\|_2 \;=\; \|\hat{g} \cdot \hat{|\psi|}\|_2
\;\leq\; \|\hat{g}\|_\infty\, \|\psi\|_2\:. \]
The function~$\hat{g}$ can be computed (ignoring factors of~$2 \pi$) to be
\[ \hat{g}(k) \;=\; c \int_0^\infty e^{-\varepsilon x}\: e^{i k x}\:dx
\;=\; \frac{c}{\varepsilon-i k} \:, \]
and thus~$\hat{g}$ is a bounded function. We conclude
that there is a constant~$C$ such that
\[ \|f\|_2 \;\leq\; C\; \|\psi\|_2\:. \]

To get a similar~$L^2$-bound on~$f'$, we first differentiate~(\ref{fdef}),
\[ f'(u) \;:=\; -\acute{\phi}(u) \grave{\phi}(u)\, \psi(u)
+ \acute{\phi}'(u) \int_u^\infty \grave{\phi}(v)\, \psi(v)\, dv \:. \]
Using~(\ref{wes}), we see that the first term is in~$L^2$. To bound the
second term, we solve the Wronskian equation~(\ref{wronski}) for~$\acute{\phi'}$
and use the above inequalities to obtain
\[ |\acute{\phi}'(u)| \;\leq\; C\, u^s \,e^{|{\mbox{\scriptsize{Im}}}\, \omega|\, u}
\qquad {\mbox{for $u \gg 0$}}\:. \]
In view of Theorem~\ref{thm31}, we have similar inequalities
for~$\acute{\phi}'$ as for~$\acute{\phi}$, and thus we can repeat the above arguments
with~$\acute{\phi}$ replaced by~$\acute{\phi}'$ to obtain
\[ \|f'\|_2 \;\leq\; C\; \|\psi\|_2\:. \]
We conclude that~$G$ is a bounded operator from~$L^2$ to~$H^{1,2}$.

It remains to show that~$G$ maps~$C^\infty_0$ into the Schwartz class.
By iteratively taking the derivatives~$(\partial_u + \partial_v)$ of~(\ref{Gdef}),
we see that~$(\partial_u + \partial_v)^n G(u,v)$ is continuous in both variables.
Since for any~$\psi \in C^\infty_0$,
\[ \partial_u \int_{-\infty}^\infty G(u,v)\: \psi(v)\: dv \;=\;
\int_{-\infty}^\infty ((\partial_u + \partial_v) G(u,v))\: \psi(v)\: dv \:+\:
\int_{-\infty}^\infty G(u,v)\: \psi'(v)\: dv \:, \]
where the last integral was obtained by partial integration,
it follows that~$G \psi$ is in~$C^1$. The higher regularity follows by induction.
To prove that~$G \psi$ has rapid decay, we choose~$u$ to the left of the support of~$\psi$. Then
\[ (G \psi)(u) \;=\; \acute{\phi}(u) \int_{-\infty}^\infty \frac{\grave{\phi}(v)\: \psi(v)}{w(\acute{\phi},
\grave{\phi})}\: dv\:. \]
Since~$\acute{\phi}$ has exponential decay, it follows that~$G \psi$ has rapid decay at~$u=-\infty$.
A similar argument using~$\grave{\phi}$ shows that~$G \psi$ has rapid decay at~$+\infty$.
Differentiating through the Schr\"odinger equation~(\ref{schroedinger}), one sees
that the derivatives of~$\acute{\phi}$ and~$\grave{\phi}$ also have rapid decay at their
respective asymptotic ends, implying that all derivatives of~$G \psi$ have rapid decay.
\QED

We now express the resolvent of~$H$ in terms of~$G$.
\begin{Thm} \label{thmresolvent}
Every complex number~$\omega$ satisfying~(\ref{oas})
lies in the resolvent set of the operator~$H$. The resolvent~$R_\omega:=(H-\omega)^{-1}$
has the integral kernel representation
\beq \label{Rdef0}
(R_\omega\, \Psi)(u) \;=\; \int_{-\infty}^\infty R_\omega(u,v)\, \Psi(v)\, dv \:,
\eeq
where
\beq \label{Rdef}
R_\omega(u,v) \;=\; \left(\! \begin{array}{cc} 0 & 0 \\ \delta(u,v) & 0 \end{array} \! \right)
\:+\: G(u,v) \left(\! \begin{array}{cc} \omega - \beta(v) & 1 \\
\omega\, (\omega - \beta(v)) & \omega \end{array} \! \right) \:,
\eeq
and~$\beta$ is defined as in~(\ref{betadef}).
\end{Thm}
{\Proof} 
A short calculation using~(\ref{Hdef}, \ref{Rdef}, \ref{Geq}) shows that
\beq \label{Rdis}
(H-\omega)\: R_\omega(u,v) \;=\; \1\, \delta(u-v)\:.
\eeq
Using Lemma~\ref{lemma42}, we can use~(\ref{Rdef0}) to define~$R_\omega$
as a bounded operator from~${\mathcal{H}}=H^{1,2} \oplus L^2$
to itself.

Let us show that the image of the operator~$(H-\omega)$ (with domain of
definition~${\mathcal{D}}(H)={\mathcal{S}}(\R)^2$)) is dense
in~${\mathcal{H}}$. To this end, for given~$\Psi
\in {\mathcal{H}}$ we choose a sequence~$\Psi_n \in C^\infty_0$
with~$\Psi_n \rightarrow \Psi$ in ${\mathcal{H}}$.
According to Lemma~\ref{lemma42}, the functions~$\Phi_n := R_\omega \Psi_n$
are Schwartz functions. Hence the~$\Phi_n$ are in the domain of~$H$,
and from~(\ref{Rdis}) we see that~$(H-\omega) \Phi_n = \Psi_n$.

We conclude that~$\omega$ lies in the resolvent set of~$H$
and that $(H-\omega)^{-1}=R_\omega$.
\QED

We end this section by showing that the boundary of the set~(\ref{oas})
lies in the essential spectrum of~$H$.
\begin{Prp} \label{prpess}
\[ \sigma_{\mbox{\scriptsize{ess}}}(H) \;\supset\; \R \cup \left(\R + \frac{is}{4M} \right) . \]
\end{Prp}
{\Proof} Let~$\omega \in \R \cup (\R+\frac{is}{4M})$ and set~$\kappa={\mbox{Re}}\, \omega$.
We choose a positive test function~$\eta \in C^\infty_0((-2,2))$
with~$\eta|_{(-1,1)} \equiv 1$ and consider for any~$L \neq 0$ the ``wave packet''
\[ \Psi_{\kappa,\omega,L}(u) \;=\; \frac{1}{L}\:\eta\!\left(\frac{u-L^3}{L^2} \right)\: e^{-i \kappa u}
\left( \!\!\begin{array}{cc} 1 \\ \omega \end{array} \!\!\right) \]
of momentum~$\kappa$, localized in the interval~$[L^3-2 L^2, L^3+2 L^2]$.
A scaling argument shows that~$\|\Psi_{\kappa, \omega, L}\|_{L^2} = \|\eta\|_{L^2}$,
and thus the Hilbert space norm $\|\Psi_{\kappa,\omega,L}\|_{\mathcal{H}}$ is
bounded away from zero as~$L \rightarrow \pm \infty$. Furthermore, moving the wave packet to infinity
and to the event horizon, respectively, we can use the asymptotic form of the Hamiltonian to obtain
\begin{eqnarray*}
\lim_{L \rightarrow \infty} \|(H-\omega)\,\Psi_{\kappa, \omega, L}\|_{\mathcal{H}} &=& 0 \spc
{\mbox{if $\omega=\kappa$}} \\
\lim_{L \rightarrow -\infty} \|(H-\omega)\,\Psi_{\kappa, \omega, L}\|_{\mathcal{H}} &=& 0 \spc
{\mbox{if $\displaystyle \omega=\kappa + \frac{is}{4M}$}}\:.
\end{eqnarray*}
Hence~$\omega$ lies in the approximate point spectrum.
\QED

\section{WKB Estimates} \label{sec4a}
\setcounter{equation}{0}
In this section we again assume that~(\ref{oas}) is satisfied and that for a suitable
constant~$K>1$ (to be determined later)
one of the following two conditions holds:
\begin{description}
\item[(C1)] $|\omega| \geq K$ and~$u \in \R$.
\item[(C2)] $\omega \neq 0$ and~$\displaystyle |u| > \frac{K}{|\omega|}$.
\end{description}
By choosing~$K$ sufficiently large, we can clearly arrange that
the potential~$V$ in~(\ref{Vdef}) has no zeros.
Then the WKB functions~$\acute{\alpha}$ and~$\grave{\alpha}$ are defined
by~(\ref{adef}). We choose the normalization
constants~$\acute{c}, \grave{c}$ such that
\[ \begin{array}{cl}
\left. \begin{array}{rcl} \displaystyle \lim_{u \rightarrow -\infty} e^{i \Omega u}\: \acute{\alpha} &=& 1 \\
\displaystyle \lim_{u \rightarrow \infty} u^s\: e^{-i \omega u}\: \grave{\alpha} &=& 1
\end{array} \right\} &  {\mbox{if $\displaystyle {\mbox{Im}}\, \omega > \frac{s}{4M}$}} , \\[2em]
\left. \begin{array}{rcl} \displaystyle \lim_{u \rightarrow -\infty} e^{-i \Omega u}\: \acute{\alpha} &=& 1 \\
\displaystyle \lim_{u \rightarrow \infty} u^{-s}\: e^{i \omega u}\: \grave{\alpha} &=& 1
\end{array} \right\} & {\mbox{if ${\mbox{Im}}\, \omega < 0 $}} .
\end{array} \]
The next theorem shows that for large~$|\omega|$ the fundamental solutions~$\acute{\phi}$
and~$\grave{\phi}$ constructed in the previous section are well-approximated by
the WKB solutions. We restrict attention to the physically interesting cases~$s=\frac{1}{2}, 1, 2$
(although the method works for arbitrary~$s$ just as well).
We let~$\rho$ be the function
\[ \rho(u)=\sqrt{1+u^2} \]
and introduce the constant~$\underline{\rho}$ by
\[ \underline{\rho} \;=\; \left\{ \begin{array}{cl} 1 & {\mbox{in case {\bf{(C1)}}}} \\
\rho(K/|\omega|) & {\mbox{in case {\bf{(C2)}}}}\:. \end{array} \right. \]
Note that in both cases, $\rho(u) > \underline{\rho}$ and~$|\omega|\underline{\rho} > K$.

\begin{Thm} \label{thm41}
Let~$s \in \{ \frac{1}{2}, 1, 2\}$. Then there are
constants~$C, K>0$ such that for all~$\omega$ and~$u$
satisfying~(\ref{oas}) and  either~{\bf{(C1)}} or~{\bf{(C2)}}, the following inequalities hold
\[ \left| \frac{\acute{\phi}}{\acute{\alpha}} - 1 \right|
\:+\: \left| \frac{\acute{\phi}'}{\acute{\alpha}'} - 1 \right| \;\leq\; \frac{4C}{|\omega|\,\underline{\rho}}
\qquad {\mbox{and}} \qquad
\left| \frac{\grave{\phi}}{\grave{\alpha}} - 1 \right|
\:+\: \left| \frac{\grave{\phi}'}{\grave{\alpha}'} - 1 \right| \;\leq\; \frac{4C}{|\omega|\,\underline{\rho}} \:. \]
\end{Thm}
{\Proof}
We only give the estimate for~$\grave{\phi}$, because~$\acute{\phi}$ can be treated similarly
after replacing~$u$ by~$-u$. By choosing~$K$ sufficiently large, we can arrange that for~$n=1,2,3$,
\begin{eqnarray}
|W(u)| &\leq& \frac{c}{\rho^3} \:,\spc \;\:
|\partial^n W(u)| \;\leq\; \frac{c}{\rho^{3+n}} \label{Wes} \\
|V(u)| &\geq& \frac{|\omega|^2}{4}\:, \spc
|\partial^n V(u)| \;\leq\; \frac{c\,|\omega|}{\rho^{1+n}} \:. \label{Ves}
\end{eqnarray}
Furthermore, it follows from our sign convention~(\ref{signconvention}) that for
all~$\omega$ satisfying~(\ref{oas}),
\[ {\mbox{Re}}\, \sqrt{V} \;\geq\; -\frac{s}{\rho} + {\mbox{O}}(\rho^{-2}) \:, \]
and integrating gives the bound
\beq \label{expV}
\left| e^{-\int_u^x 2 \sqrt{V}} \right| \;\leq\; \tilde{c} \left(\frac{\rho(x)}{\rho(u)} \right)^{2s}
\spc {\mbox{for all~$x \geq u$}}.
\eeq
Using~(\ref{adef}, \ref{grdef}, \ref{pert2}), the functions~$E^{(l)}$ defined by
\beq \label{Eldef}
E^{(l)}(u) \;=\; \frac{\phi^{(l)}(u)}{\grave{\alpha}(u)}
\eeq
satisfy the relations
\beq  \label{Eind}
\left. \begin{array}{rcl} E^{(0)} &\equiv& 1 \\[.5em]
E^{(l+1)}(u) &=& \displaystyle
 \int_u^\infty \frac{W(x)}{2\, \sqrt{V(x)}}
\left\{ 1 - e^{-2 \int_u^x \sqrt{V}} \right\} E^{(l)}(x)\: dx \:.
\end{array} \right\}
\eeq

We begin with the case~$s=1$. We shall prove inductively that for sufficiently
large~$C$ there are constants~$a^{(l)}$ and~$b^{(l)}$ such that for all~$l \geq 0$
the following inequalities hold,
\beq \label{se1}
\left| E^{(l)}(u) - a^{(l)} \right| \;\leq\; \frac{b^{(l)}}{\rho(u)}
\qquad {\mbox{with}} \qquad
|a^{(l)}|+|b^{(l)}| \;\leq\; \left( \frac{C}{|\omega| \, \underline{\rho}} \right)^l \:.
\eeq
To satisfy these conditions in the case~$l=0$, we simply set~$a^{(0)}=1$ and~$b^{(0)}=0$.
Thus assume that~(\ref{se1}) holds for a given~$l$. Since~$\rho \geq 1$, (\ref{se1}) implies that
\beq \label{se2}
|E^{(l)}(u)| \;\leq\; |a^{(l)}| + |b^{(l)}| \:.
\eeq
The estimates
\begin{eqnarray*}
\left| \int_u^\infty \frac{W(x)}{2\, \sqrt{V(x)}}\: E^{(l)}(x)\: dx \right|
&\leq& \frac{1}{|\omega|}
\int_u^\infty \frac{c}{\rho(x)^3}\: \left( \frac{C}{|\omega| \, \underline{\rho}}
\right)^l dx \\
&\leq& \frac{c\, C^l}{(|\omega| \underline{\rho})^{l+1}} 
\int_{-\infty}^\infty \frac{1}{1+x^2}\: dx \;=\;
\frac{c\pi\, C^l}{(|\omega| \underline{\rho})^{l+1}}
\end{eqnarray*}
give us control of the first term in the curly brackets in~(\ref{Eind}).
To estimate the second term in the curly brackets, we first consider the
error term in~(\ref{se1}),
\begin{eqnarray*}
\left| \int_u^\infty \frac{W(x)}{2\, \sqrt{V(x)}}\:e^{-2 \int_u^x \sqrt{V}}\: \frac{b^{(l)}}{\rho(x)}\: dx \right|
&\leq& \frac{1}{|\omega|}
\int_u^\infty \frac{c}{\rho(x)^3}
\:\tilde{c} \left(\frac{\rho(x)}{\rho(u)} \right)^2\:\frac{1}{\rho(x)}\: \frac{C^l}{(|\omega|
\underline{\rho})^l}\:  dx \\
&\leq& \frac{C^l c\tilde{c}}{(|\omega| \underline{\rho})^{l+1}}
\int_u^\infty \frac{1}{\rho(x)^2}\:  dx \;\leq\;
\frac{C^l \pi c\tilde{c}}{(|\omega| \underline{\rho})^{l+1}}\:.
\end{eqnarray*}
For the constant term in~(\ref{se1}) we can integrate by parts
\begin{eqnarray}
\lefteqn{ \int_u^\infty \frac{W(x)}{2\, \sqrt{V(x)}}\:e^{-2 \int_u^x \sqrt{V}}\: a^{(l)}\: dx
\;=\; -a^{(l)} \int_u^\infty \frac{W(x)}{4\, V(x)}\:
\frac{d}{dx} \left( e^{-2 \int_u^x \sqrt{V}} \right)\: dx } \nonumber \\
&=& a^{(l)} \frac{W(u)}{4\, V(u)}
\:+\: a^{(l)} \int_u^\infty 
\left(\frac{W(x)}{4\, V(x)} \right)'\:
e^{-2 \int_u^x \sqrt{V}} \: dx \spc\spc
\label{ibp}
\end{eqnarray}
to get
\begin{eqnarray*}
\lefteqn{ \left| \int_u^\infty \frac{W(x)}{2\, \sqrt{V(x)}}\:e^{-2 \int_u^x \sqrt{V}}\: a^{(l)}\: dx \right| } \\
&\leq& \frac{C^l}{2 |\omega|^{l+2}\, \underline{\rho}^l}\:\frac{c}{\rho(u)^3}
\:+\: \frac{C^l}{(|\omega| \underline{\rho})^l} \int_u^\infty
\frac{c \tilde{c}}{|\omega|^2\, \rho(x)^4} \left(\frac{\rho(x)}{\rho(u)} \right)^2 dx \\
&=& \frac{C^l}{2 |\omega|^{l+2}\, \underline{\rho}^l}\:\frac{c}{\rho(u)^3}
\:+\: \frac{C^l c \tilde{c} \pi}{|\omega|^{l+2}\, \underline{\rho}^l\, \rho(u)^2}
\;\leq\; \frac{C^l c}{2 (|\omega| \underline{\rho})^{l+1}}
\:+\: \frac{C^l c \tilde{c} \pi}{(|\omega| \underline{\rho})^{l+1}} \:.
\end{eqnarray*}
Choosing~$C>c \pi + c \tilde{c} \pi + (c/2 +\pi c\tilde{c})$, the induction
step is thereby complete, so that~(\ref{se1}) holds.

We choose~$K>4 C$. Then the inequality~$|\omega|\underline{\rho} >K$ implies
that~$K/(|\omega| \underline{\rho}) < \frac{1}{4}$, and thus 
\begin{eqnarray*}
\left| \frac{\grave{\phi}}{\grave{\alpha}} -1 \right| &\stackrel{(\ref{Eldef})}{=}&
\left| \left( \sum_{l=0}^\infty E^{(l)} \right) - 1 \right| 
\;\stackrel{(\ref{Eind})}{\leq}\; \sum_{l=1}^\infty |E^{(l)}|
\;\stackrel{(\ref{se2})}{\leq}\; \sum_{l=1}^\infty \left( |a^{(l)}| + |b^{(l)}| \right) \\
&\stackrel{(\ref{se1})}{\leq}& \sum_{l=1}^\infty \left( \frac{C}{|\omega|\, \underline{\rho}}
\right)^l
\;=\; \frac{C}{|\omega| \underline{\rho} - C} \;\leq\; \frac{4C}{3|\omega| \underline{\rho}}\:.
\end{eqnarray*}
To treat the derivative of~$\grave{\phi}$, we first note that
\[ \frac{(\phi^{(l)})'}{\grave{\alpha}'} \;=\; (E^{(l)})'\: \frac{\grave{\alpha}}{\grave{\alpha}'}
+ E^{(l)}\:. \]
Differentiating~(\ref{Eind}) and using~(\ref{adef}), we find that
\[ (E^{(l+1)})'\: \frac{\grave{\alpha}}{\grave{\alpha}'}
\;=\; -\left[ 1 + \frac{V'}{4 V^{\frac{3}{2}}} \right]^{-1}
 \int_u^\infty \frac{W(x)}{\sqrt{V(x)}}\:
e^{-2 \int_u^x \sqrt{V}} \: E^{(l)}(x)\: dx\:. \]
Using~(\ref{Ves}), one sees that in both cases~{\bf{(C1)}} and~{\bf{(C2)}},
the square bracket is uniformly bounded away from zero,
and the integral can be estimated exactly as before. This shows that
\[ \sum_{l=1}^\infty \left|(E^{(l+1)})'\: \frac{\grave{\alpha}}{\grave{\alpha}'} \right| \]
can be made arbitrarily small by choosing~$K$ large enough.
Thus
\[ \left| \frac{\acute{\phi}'}{\acute{\alpha}'} - 1 \right| \;\leq\; \frac{4C}{3 |\omega| \underline{\rho}} \:, \]
and this concludes the proof of the theorem in the case~$s=1$.

In the case~$s=\frac{1}{2}$ the proof is even easier since we do not need
to integrate by parts in~(\ref{ibp}).
Finally, if~$s=2$, we again consider solutions of~(\ref{Eind}), but we replace~(\ref{se1}) by
\beq \label{se3}
\left| E^{(l)}(u) - a^{(l)} - \frac{b^{(l)}}{\rho(u)}
- \frac{c^{(l)}}{\rho(u)^2} \right| \;\leq\; \frac{d^{(l)}}{\rho(u)^3}
\eeq
and our task is to prove inductively that there are constants~$C$  and~$K$ such that
for all~$l$ and~$\omega$ satisfying the conditions~{\bf{(C1)}} or~{\bf{(C2)}},
\beq \label{se4}
|a^{(l)}| + |b^{(l)}| + |c^{(l)}| + |d^{(l)}| \;\leq\;
\left( \frac{C}{|\omega| \, \underline{\rho}} \right)^l \:.
\eeq
Once these inequalities are proved, the theorem follows as in the case~$s=1$.
Again for~$l=0$, setting~$a^{(0)}=1$ and~$b^{(0)}=c^{(0)}=d^{(0)}=0$,
there is nothing to prove. The induction step follows as in the
case~$s=1$, however we here need to integrate by parts up to three times.
We shall not give all the details, but merely consider the term involving~$b^{(l)}$,
which is representative of all other terms. After integrating by parts twice, we get
\begin{eqnarray*}
\lefteqn{ \int_u^\infty \frac{W(x)}{2\, \sqrt{V(x)}}\:e^{-2 \int_u^x \sqrt{V}}\: \frac{b^{(l)}}{x}\: dx } \\
&=& b^{(l)} \frac{W(u)}{4\, V(u)\: u}
\:+\: b^{(l)} \int_u^\infty  \left(\frac{W(x)}{4\, V(x)\:x} \right)'\:
e^{-2 \int_u^x \sqrt{V}} \: dx \\
&=& b^{(l)} \frac{W(u)}{4\, V(u)\: u}
\:+\: \frac{b^{(l)}}{2 \sqrt{V(u)}}\: \left(\frac{W(x)}{4\, V(x)\:x} \right)' \\
&&+ b^{(l)} \int_u^\infty 
\left( \frac{1}{2 \sqrt{V(x)}}
\left(\frac{W(x)}{4\, V(x)\:x} \right)' \right)'\:
e^{-2 \int_u^x \sqrt{V}} \: dx\:.
\end{eqnarray*}
Carrying out the differentiations, we can take absolute values
and estimate term by term using~(\ref{Wes}, \ref{Ves}).
Possibly after increasing~$K$, we get the desired result.
\QED

\section{Resolvent Estimates for Large~$|\omega|$} \label{secres}
\setcounter{equation}{0}
In this section we assume again that~$\omega$ is in the range~(\ref{oas})
and that condition~{\bf{(C1)}} holds,
so that the WKB estimates of Theorem~\ref{thm41} are valid.
Our goal is to prove the following estimate of the resolvent for large~$|\omega|$,
which will play a crucial role in the completeness proof.
\begin{Thm} \label{thm51}
For every~$\Psi \in C^\infty_0(\R)^2$ and~$u \in \R$
there are constants~$K, C=C(u)>0$ such that
for all~$\omega$ satisfying~(\ref{oas}),
\[ |(R_\omega \Psi)(u)| \;\leq\; \frac{C}{|\omega|}\:. \]
\end{Thm}
{\Proof}
Noting that for any~$\psi \in C^\infty_0(\R)$,
\[ \int_{-\infty}^\infty \delta(u-v)\, \psi(v)\, dv \;=\;
\int_{-\infty}^\infty G(u,v) \left(-\partial_v^2 + V \right) \psi(v) \:, \]
a short calculation using~(\ref{Rdef0}, \ref{Rdef}) allows us to write
the resolvent for any~$\Psi \in C^\infty_0(\R)^2$ as 
\[ (R_\omega \Psi)(u) \;=\; \int_{-\infty}^\infty G(u,v)\, N \Psi
\qquad {\mbox{where}} \qquad
N \;=\; \left( \!\begin{array}{cc} \omega - \beta & 1 \\ \alpha & \omega
\end{array} \!\right) \:. \]
Thus, since~$N$ is linear in~$\omega$, the result will hold if
we show that for every~$\psi \in C^\infty_0(\R)$,
\beq \label{Gbound}
|(G \psi)(u)| \;\leq\; \frac{C}{|\omega|^2} \:.
\eeq

Before giving the proof of~(\ref{Gbound}), we
collect a few properties of the potential~$V$ for large~$|\omega|$.
It is obvious from~(\ref{Vdef}) that there is a constant~$m>0$ such that
for all~$\omega$ satisfying~(\ref{oas}) and~$u \in \R$,
\beq \label{Vestimate}
|V(u)| \;\geq\; \frac{|\omega|^2}{4}\:,\spc
|V'(u)| \;\leq\; m\, |\omega| \:.
\eeq
Furthermore, writing the potential in the form
\[ V \;=\; -\omega^2 + \omega \beta + f \]
(where~$\beta$ and~$f$ are independent of~$\omega$),
its square root can be written as
\[ \sqrt{V(u)} \;=\; \pm i \omega \: \sqrt{ 1 - \frac{\beta(u)}{\omega} - \frac{f(u)}{\omega^2} }
\:. \]
Using our sign convention~(\ref{signconvention}) together with the fact
that~$\beta$ and~$f$ are bounded functions,
we conclude that there is a constant~$\tilde{m}>0$ such that
for all~$\omega$ satisfying~(\ref{oas}) and~$u \in \R$,
\beq \label{sqVest}
{\mbox{Re}}\, \sqrt{V(u)} \;\geq\; -\tilde{m} \:.
\eeq

Using~(\ref{Gdef}), we have
\beq \label{Geq2}
(G \psi)(u) \;=\; \frac{1}{w(\acute{\phi}, \grave{\phi})}
\left( \grave{\phi}(u) \int_{-\infty}^u \acute{\phi}(v)\, \psi(v) \:dv
\:+\: \acute{\phi}(u) \int_u^\infty \grave{\phi}(v)\, \psi(v) \:dv \right) .
\eeq
We first estimate the Wronskian. From~(\ref{adef}), we have
\begin{eqnarray}
\acute{\alpha} \grave{\alpha} &=& \frac{\acute{c} \grave{c}}{\sqrt{V}} \\
\acute{\alpha}' &=& \left( -\frac{V'}{4V} + \sqrt{V} \right) \acute{\alpha}\:,\qquad
\grave{\alpha}' \;=\; \left( -\frac{V'}{4V} - \sqrt{V} \right) \grave{\alpha}\:,
\label{apeq}
\end{eqnarray}
so
\begin{eqnarray}
w(\acute{\phi}, \grave{\phi}) &=&
\frac{\acute{\phi}'}{\acute{\alpha}'}\, \frac{\grave{\phi}}{\grave{\alpha}}\:
\acute{\alpha}' \, \grave{\alpha} \:-\:
\frac{\acute{\phi}}{\acute{\alpha}}\, \frac{\grave{\phi}'}{\grave{\alpha}'}\:
\acute{\alpha} \, \grave{\alpha}' \\
&=& \acute{c} \grave{c}\: \frac{\acute{\phi}'}{\acute{\alpha}'}\, \frac{\grave{\phi}}{\grave{\alpha}}
\left( -\frac{V'}{4V^{\frac{3}{2}}} + 1 \right)
\:-\: \acute{c} \grave{c}\:\frac{\acute{\phi}}{\acute{\alpha}}\, \frac{\grave{\phi}'}{\grave{\alpha}'}
\left( -\frac{V'}{4V^{\frac{3}{2}}} - 1 \right) .
\end{eqnarray}
Applying Theorem~\ref{thm41}, we obtain, possibly after increasing~$C$,
\beq \label{westimate}
\left| \frac{w(\acute{\phi}, \grave{\phi})}{2 \acute{c} \grave{c}} - 1 \right|
\;\leq\; \frac{C}{|\omega|}\:.
\eeq

Next we consider the second term in the brackets in~(\ref{Geq2}). We have
\begin{eqnarray}
\lefteqn{ \int_u^\infty \grave{\phi}(v)\, \psi(v) \:dv \;=\;
\int_u^\infty \grave{\phi}'
\:\frac{\grave{\phi}\, \psi}{\grave{\phi}'}\:dv \;=\;
\left.
\:\frac{\grave{\phi}^2\, \psi}{\grave{\phi}'}
\right|_{u} \:-\:
\int_u^\infty \grave{\phi}\,
\left( \frac{\grave{\phi}\, \psi}{\grave{\phi}'}
\right)' dv } \nonumber \\
&=& \left.
\:\frac{\grave{\phi}^2\, \psi}{\grave{\phi}'}
\right|_{u} \:-\: \int_u^\infty
\frac{\grave{\phi}}{\grave{\phi}'}\: \grave{\phi}\, \psi'
\:+\: \int_u^\infty
\left( \frac{\grave{\phi}''\, \grave{\phi}}{\grave{\phi}'^2} - 1 \right)
\grave{\phi}\, \psi\:.
 \label{pint2}
\end{eqnarray}
Squaring the identity
\[ \frac{\grave{\phi}'}{\grave{\phi}} \;\stackrel{(\ref{apeq})}{=}\; \frac{\grave{\phi}'}{\grave{\alpha}'}\:
\frac{\grave{\alpha}}{\grave{\phi}} \left( -\frac{V'}{4V} - \sqrt{V} \right) \]
and using the equation~$\grave{\phi}'' = V \grave{\phi}$, we find that
\[ \frac{\grave{\phi}'' \grave{\phi}}{\grave{\phi}'^2} \;=\;
V\: \frac{\grave{\phi}^2}{\grave{\phi}'^2} \;=\;
\left( \frac{\grave{\alpha}'}{\grave{\phi}'}\:
\frac{\grave{\phi}}{\grave{\alpha}} \right)^2 \left(1 +\frac{V'}{2V^{\frac{3}{2}}} +
\frac{V'^2}{16 V^3} \right)^{-1}\:. \]
Hence
\beq \label{exp}
\left( \frac{\grave{\phi}''\, \grave{\phi}}{\grave{\phi}'^2} - 1 \right)
\;=\; \left(
\left[ \left( \frac{\grave{\alpha}'}{\grave{\phi}'}\:
\frac{\grave{\phi}}{\grave{\alpha}} \right)^2 -1 \right] +\frac{V'}{2V^{\frac{3}{2}}} -
\frac{V'^2}{16 V^3} \right) \left(1 +\frac{V'}{2V^{\frac{3}{2}}} +
\frac{V'^2}{16 V^3} \right)^{-1} ,
\eeq
and using Theorem~\ref{thm41} together with~(\ref{Vestimate}), we see that~(\ref{exp})
is of order~$1/|\omega|$.
Now we can estimate~(\ref{pint2}) termwise to obtain
\[ \left| \int_u^\infty \grave{\phi}(v)\, \psi(v)\, dv \right| \;\leq\;
\frac{c(\psi)}{|\omega|}\: \sup_{\mathcal{K}} |\grave{\phi}|\:, \]
where~${\mathcal{K}}$ denotes the support of~$\psi$.

The first term in the brackets in~(\ref{Geq2}) can be estimated in a similar way. We thus
obtain
\[ |(G \psi)(u)| \;\leq\; \frac{c(\psi)}{|\omega|} \;\sup_{v \in {\mathcal{K}}}
\frac{1}{|w(\acute{\phi}, \grave{\phi})|}
\left(|\acute{\phi}(v)\, \grave{\phi}(u)| \: \Theta(u-v) + 
|\acute{\phi}(u)\, \grave{\phi}(v)|\: \Theta(v-u) \right) \:. \]
Now from~(\ref{adef}) and~(\ref{sqVest}), we get in the case~$v \geq u$,
\[ |\acute{\phi}(u)\, \grave{\phi}(v)| \;=\;
|\acute{c} \grave{c}|\: |V(u)\, V(v)|^{-\frac{1}{4}}
e^{-\int_u^v {\mbox{\scriptsize{Re}}}\, \sqrt{V}} \;\leq\;
|\acute{c} \grave{c}|\: |V(u)\, V(v)|^{-\frac{1}{4}}
e^{\tilde{m}\:(v-u)} \;\leq\; \frac{C\, |\acute{c} \grave{c}|}{|\omega|}\:, \]
where~$C$ clearly depends on~${\mathcal{K}}$. The case~$u>v$ can be treated
in a similar way. We finally apply~(\ref{westimate}) and possibly increase~$K$
to obtain~(\ref{Gbound}).
\QED

\section{An Integral Representation of the Propagator} \label{seccom}
\setcounter{equation}{0}
In this section we shall express a given~$\Psi \in C^\infty_0(\R)^2$
in terms of a contour integral of the resolvent.
Our method avoids spectral theory and Hilbert space techniques. Instead, it 
uses an idea which we learned from A.\ Bachelot~\cite[Proof of Theorem~2.12]{B} and is based upon the
resolvent estimates of Theorem~\ref{thm51}.
The result in this section is in preparation for the integral representation
of the propagator which will be derived in Section~\ref{secdecay}.

\begin{figure}[tbp]
\begin{center}
\input{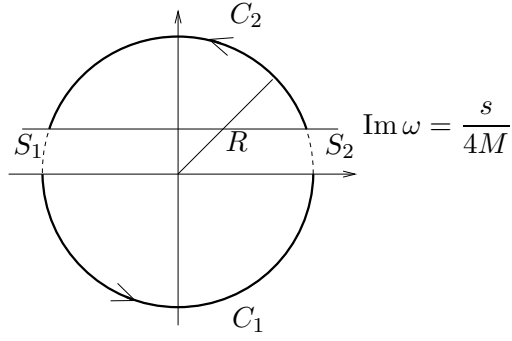}
\caption{The contour~$C_R= C_1 \cup C_2$.}
\label{fig1}
\end{center}
\end{figure}

For given~$R>0$ we consider the two contours~$C_1$ and~$C_2$ in the complex
$\omega$-plane defined by
\[ C_1 \;=\; \partial B_R(0) \cap \{ {\mbox{Im}}\, \omega < 0 \} \:,\spc
C_2 \;=\; \partial B_R(0) \cap \left\{ {\mbox{Im}}\, \omega > \frac{s}{4M} \right\} , \]
both taken with positive orientation; see Figure~\ref{fig1}.
 We set~$C_R = C_1 \cup C_2$.
We can now state the following completeness result, valid for any spin~$s \in \{\frac{1}{2}, 1, 2\}$.
\begin{Thm} \label{thmcomplete}
For every~$\Psi \in C^\infty_0(\R)^2$ and~$u \in \R$, we have the representation
\beq \label{psirep}
\Psi(u) \;=\; -\frac{1}{2 \pi i} \lim_{R \rightarrow \infty}
\int_{C_R} (R_\omega \Psi)(u)\: d\omega\:.
\eeq
\end{Thm}
{\Proof} Since the length of the contour~$S_1 \cup S_2 := \partial B_R(0) \setminus C$
stays bounded for large~$R$ (see Figure~\ref{fig1}),
\[ \left| \oint_{\partial B_R(0)} \frac{d\omega}{\omega} -
\int_{C_R}  \frac{d\omega}{\omega} \right| \;\leq\; 
\frac{1}{R} \int_{S_1 \cup S_2} |d\omega| 
\;\stackrel{R \rightarrow \infty}{\longrightarrow}\; 0\:. \]
As a consequence,
\beq \label{contour}
\frac{1}{2 \pi i} \lim_{R \rightarrow \infty} \int_{C_R} \frac{d\omega}{\omega}
\;=\; 1\:.
\eeq

Since our contours (omitting the end points) lie in the resolvent set
of~$H$ (see Theorem~\ref{thmresolvent}), we know that for every~$\omega
\in C_R$,
\[ \Psi \;=\; R_\omega\, (H-\omega) \Psi\:. \]
Dividing by~$\omega$ and integrating over~$C_R$, we can apply~(\ref{contour})
to obtain
\begin{eqnarray*}
\Psi(u) &=& \frac{1}{2 \pi i} \lim_{R \rightarrow \infty}
\int_{C_R} \frac{d\omega}{\omega}\: (R_\omega\, (H-\omega) \Psi)(u)\:. \\
&=& -\frac{1}{2 \pi i} \lim_{R \rightarrow \infty}
\int_{C_R} \left\{ (R_\omega \Psi)(u) \:-\:
\frac{1}{\omega} (R_\omega\, H \Psi)(u) \right\} d\omega\:.
\end{eqnarray*}
But the second term in the curly brackets vanishes in the limit, because using~Theorem~\ref{thm51}
and the fact that~$H \Psi \in C^\infty_0$, we have
\[ \left| \int_{C_R} (R_\omega\, H \Psi)(u) \frac{d\omega}{\omega} \right|
\;\leq\; \int_{C_R} \frac{C}{|\omega|}\: \frac{|d\omega|}{|\omega|} \;\leq\; \frac{2 \pi C}{R}\;. \]
Thus~(\ref{psirep}) holds.
\QED

Our next objective is to derive an integral representation for the solution of the Cauchy problem.
We consider the solution~$\Psi(t,u)=(\phi, i \partial_t \phi)$ of the separated Teukolsky equation~(\ref{teq2}) for initial data~$\Psi_0 \in C^\infty_0(\R)^2$. The difficulty is that~$(R_\omega \Psi_0)(u)$ only
decays in~$\omega$ like~$1/\omega$ and thus we cannot take the limit~$R \rightarrow \infty$
in~(\ref{psirep}) using the Lebesgue dominated convergence theorem, nor can we commute differentiation
with the limit. To remedy this, we derive a finite Laurent expansion of~$(R_\omega \Psi_0)(u)$.
\begin{Lemma} \label{Laurent}
For every~$n \in \N$ and~$\omega$ in the resolvent set of~$H$,
\beq \label{Rident}
(R_\omega \Psi_0)(u) \;=\; -\frac{\Psi_0(u)}{\omega} - \frac{(H \Psi_0)(u)}{\omega^2}
- \cdots - \frac{(H^n \Psi_0)(u)}{\omega^n} + \frac{1}{\omega^n}\: (R_\omega H^n \Psi_0)(u)\:.
\eeq
In particular, for~$n=3$ we have for large~$|\omega|$ the expansion
\begin{eqnarray} \label{Rident2}
(R_\omega \Psi_0)(u) &=& -\frac{\Psi_0(u)}{\omega+i} - \frac{i \Psi_0(u) + (H \Psi_0)(u)}{(\omega+i)^2} \nonumber \\
&&\qquad\quad\;\;\; - \frac{-\Psi_0(u) + 2i (H \Psi_0)(u) + (H^2 \Psi_0)(u)}{(\omega+i)^3}
\:+\: {\mathcal{O}}\!\left(\frac{1}{|\omega|^{4}}\right).
\end{eqnarray}
\end{Lemma}
{\Proof} Dividing the equation~$R_\omega(H-\omega)=\1$ by~$\omega$ gives
\[ R_\omega \Psi_0 \;=\; -\frac{\Psi_0}{\omega} + \frac{1}{\omega}\: R_\omega H \Psi_0 \:, \]
and since~$H \Psi_0$ is again in~$C^\infty_0$, we can iterate this formula to get~(\ref{Rident}).
The equation~(\ref{Rident2}) follows from~(\ref{Rident}) using the Taylor expansions
\begin{eqnarray*}
\frac{1}{\omega} &=& \frac{1}{\omega+i} \: \frac{1}{\displaystyle \left( 1 - \frac{i}{\omega+i} \right)}
\;=\; \frac{1}{\omega+i} + \frac{i}{(\omega+i)^2} - \frac{1}{(\omega+i)^3} + {\mathcal{O}}(|\omega|^{-4}) \\
\frac{1}{\omega^2} &=& \frac{1}{(\omega+i)^2} + \frac{2i}{(\omega+i)^3} + {\mathcal{O}}(|\omega|^{-4})\:.
\end{eqnarray*}

\vspace*{-.7cm}
\QED

\begin{figure}[tbp]
\begin{center}
\input{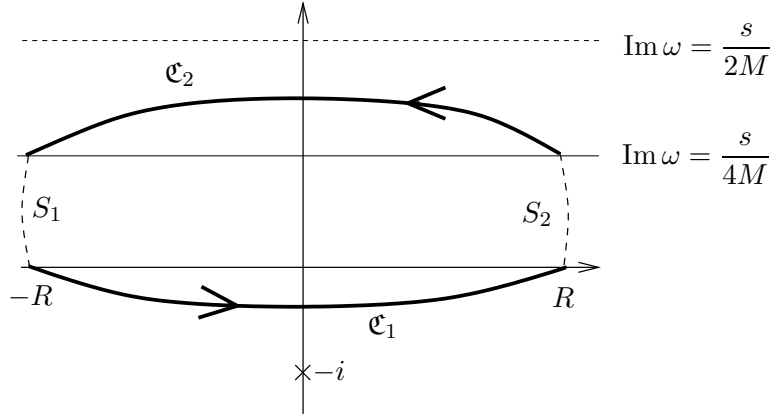}
\caption{The contour~$\mathfrak{C}_R = {\mathfrak{C}}_1 \cup {\mathfrak{C}}_2$.}
\label{fig2}
\end{center}
\end{figure}

For the application to the Cauchy problem, it is more convenient to deform the contour~$C_R$ to
the contour~${\mathfrak{C}}_R = {\mathfrak{C}}_1 \cup {\mathfrak{C}}_2$ as shown in Figure~\ref{fig2}.
This contour deformation is immediately justified from the analyticity of the resolvent in the respective
regions.
\begin{Thm} \label{thmintrep1}
For any spin $s \in \{\frac{1}{2}, 1, 2\}$,
the solution of the Cauchy problem for the separated Teukolsky equation~(\ref{teq2})
with initial data~$\Psi_0 = (\phi, \partial_t \phi)|_{t=0} \in C^\infty_0(\R)^2$
has the following representation:
\beq \label{intrep}
\Psi(t,u) \;=\; -\frac{1}{2 \pi i} \lim_{R \rightarrow \infty} \int_{{\mathfrak{C}}_R} e^{-i \omega t}\:
(R_\omega \Psi_0)(u)\: d\omega\:.
\eeq
\end{Thm}
{\Proof} Let us first verify that the limit~$R \rightarrow \infty$ in~(\ref{intrep}) exists. To this end,
we first note that inserting the ``counter terms''~$c_k/(\omega+i)^k$ does not change the integral
in the limit~$R \rightarrow \infty$,
\beq \label{counter}
\lim_{R \rightarrow \infty} \int_{{\mathfrak{C}}_R} e^{-i \omega t}\:
(R_\omega \Psi_0)(u)\: d\omega \;=\;
\lim_{R \rightarrow \infty} \int_{{\mathfrak{C}}_R} e^{-i \omega t}\:
\left((R_\omega \Psi_0)(u) - \sum_{k=1}^3 \frac{c_k(u)}{(\omega+i)^k} \right) d\omega \:.
\eeq
This is easily verified using the Cauchy integral formula for the closed contour~${\mathfrak{C}}_R \cup S_1 \cup S_2$
together with the fact that the integral over the
contours~$S_1$ and~$S_2$ vanishes as~$R \rightarrow \infty$ due to the ${\mathcal{O}}(|\omega|^{-1})$-decay of the
counter terms.
By choosing the coefficients~$c_k$ as in~(\ref{Rident2}), we can arrange that
the integrand decays like~$|\omega|^{-4}$. Thus we can apply
Lebesgue's dominated convergence theorem to see that the limit~$R \rightarrow \infty$ exists.

Setting~$t=0$, it follows from Theorem~\ref{thmcomplete} that~$\Psi$
satisfies the correct initial conditions.
To see that~$\Psi$ is a solution of the Teukolsky equation, we apply the operator~$(i \partial_t - H)$
to~(\ref{counter}).
Since taking the time derivative of the integrand gives a factor of~$-i \omega$, whereas
the WKB-estimates of Theorem~\ref{thm41} show that the spatial derivatives of~$(R_\omega \Psi_0)(u)$
scale like powers of~$\omega$, we see that the partial derivatives of the integrand
on the right side of~(\ref{counter}) can all be dominated by a constant times~$|\omega|^{-2}$.
Hence we can interchange the differentiations with the limit and the integration to obtain
\begin{eqnarray*}
\lefteqn{ (i\partial_t - H) \lim_{R \rightarrow \infty} \int_{{\mathfrak{C}}_R} e^{-i \omega t}\:
(R_\omega \Psi_0)(u)\:d\omega } \\
&=& -\lim_{R \rightarrow \infty} \int_{{\mathfrak{C}}_R} e^{-i \omega t}\:
(H-\omega) \left((R_\omega \Psi_0)(u) - \sum_{k=1}^3 \frac{c_k(u)}{(\omega+i)^k} \right) d\omega\:.
\end{eqnarray*} 
If the operator~$H$ acts on the factors~$c_k(u)$, the resulting expressions are exactly
of the form as considered after~(\ref{counter}) and vanish in the limit.
If~$\omega$ multiplies~$c_2$ or~$c_3$, the resulting terms again vanish in the limit.
Hence, using that~$c_1 = -\Psi_0(u)$, we obtain
\begin{eqnarray*}
\lefteqn{ (i\partial_t - H) \lim_{R \rightarrow \infty} \int_{{\mathfrak{C}}_R} e^{-i \omega t}\:
(R_\omega \Psi_0)(u)\:d\omega } \\
&=& -\lim_{R \rightarrow \infty} \int_{{\mathfrak{C}}_R} \!\! e^{-i \omega t}\:
\left( \Psi_0(u) - \omega \frac{\Psi_0(u)}{(\omega+i)} \right) d\omega
\;=\; -\lim_{R \rightarrow \infty} \int_{{\mathfrak{C}}_R} \!\! e^{-i \omega t}\:
\left( \frac{i \Psi_0(u)}{(\omega+i)} \right) d\omega \;=\; 0 \:,
\end{eqnarray*}
where we again used the argument after~(\ref{counter}).
\QED

\section{Contour Deformations Onto The Real Line} \label{seccontour}
\setcounter{equation}{0}
The objective of this section is to move that part of the contour, which in Theorem~\ref{thmcomplete} lies in the lower half plane, onto the real line.
Since by Theorem~\ref{thm31}, $\acute{\phi}$ is holomorphic in a neighborhood
of the real line, our task is to analyze for any given~$\omega_0 \in \R$
the Jost solutions~$\grave{\phi}$ for~$\omega$ in the the set
\[ C_\varepsilon(\omega_0) \;=\; \left\{ |\omega-\omega_0| < \varepsilon {\mbox{ and }}
{\mbox{Im}} \,\omega < 0 \right\} \]
and consider their limiting behavior as~$\omega \rightarrow \omega_0$.
We distinguish the cases~$\omega_0 =0$ and~$\omega_0 \neq 0$.

We begin with the first case~$\omega_0=0$.
Qualitatively, if~$u \gg |\omega|^{-1}$, the solution~$\grave{\phi}$
is well-approximated by the WKB solution (see Theorem~\ref{thm41}).
If on the other hand~$u \ll |\omega|^{-1}$, the solution should be close to the
solution of the Schr\"odinger equation for~$\omega=0$. In order to match the asymptotics
through the intermediate region, we need a separate argument based on classical Whittaker functions.

\begin{Lemma}
Let~$\grave{\phi}$ be the Jost functions constructed in Theorem~\ref{thm34} for~$\omega \in C_\varepsilon(0)$.
After suitable rescaling, the limit~$\omega \rightarrow 0$ exists,
\beq \label{philim}
\lim_{C_\varepsilon(0) \ni \omega \rightarrow 0} \omega^{s+\sigma} \grave{\phi} \;=\; \grave{\phi}_0\:,
\eeq
where
\beq \label{sigmadef}
\sigma \;=\; \frac{1}{2} \left(\sqrt{1+4 s^2 + 4 \lambda} - 1 \right) .
\eeq
The limit function~$\grave{\phi}_0$ is a solution of the Schr\"odinger equation~(\ref{schroedinger}) for~$\omega=0$
and has the asymptotics
\beq \label{phi0asy}
\lim_{r \rightarrow \infty} \left( r^\sigma\; \grave{\phi}_0 \right) \;=\; 
\frac{(-4)^{-\frac{\sigma}{4}} \,\Gamma(2 \sigma+2)}{(2i)^s \,\Gamma(\sigma + 1-s)}\;.
\eeq
\end{Lemma}
{\Proof} In order to avoid the difficulties associated with the term
proportional to~$u^{-2} \log u$ in the potential~(\ref{Vasy}), it is easier to work in
the coordinate~$r$. Introducing the function
\beq \label{psieq}
\psi(r) \;=\; \sqrt{\Delta}\, R(r) \;=\; \frac{\sqrt{\Delta}}{r} \, \phi(r)\:,
\eeq
we can write the Schr\"odinger equation~(\ref{schroedinger}) as
\beq \label{schV}
-\frac{d^2}{dr^2} \psi(r) + {\mathcal{V}}(r)\, \psi(r) \;=\; 0\:,
\eeq
where the new potential~${\mathcal{V}}$ has the following asymptotics near infinity:
\beq \label{calV}
{\mathcal{V}}(r) \;=\; - \omega^2 - 2\, \frac{i s \omega + M \omega^2}{r} +
\frac{s^2 + \lambda - 2 i M s \omega - 12 M^2 \, \omega^2}{r^2}\:+\: {\mathcal{O}}(r^{-3})\: .
\eeq

We first consider the equation~(\ref{schV}) where we simply drop the error
term in~(\ref{calV}). Then this modified equation can be solved
exactly using Whittaker functions~\cite[Chapter~13, pp 505-508]{AS}.
The two fundamental solutions are~$M_{\kappa, \mu}(z)$ and~$W_{\kappa, \mu}(z)$,
where the parameters are given by
\[ \kappa \;=\; s - 2 i \omega M\:,\quad
\mu \;=\; \frac{1}{2} \, \sqrt{1 + 4 s^2 + 4 \lambda - 8 i M s \omega - 48\, M^2 \omega^2}\:,\quad
z \;=\; 2 i \omega r\:. \]
The function~$\grave{\phi}$ clearly is a linear combination of~$M_{\kappa, \mu}(z)$ and~$W_{\kappa, \mu}(z)$.
Comparing the asymptotics for large~$|z|$~\cite[(13.5.1) and (13.5.2)]{AS} with the
asymptotics of~$\grave{\phi}$ (\ref{phiasy}), we can determine the coefficients of this
linear combination to obtain for the function~$\grave{\psi} = \sqrt{\Delta} \grave{\phi}/r$
\[ \grave{\psi} \;=\; (2 i \omega)^{-s + 2 i M \omega} \:W_{\kappa, \mu}(z) \:. \]
Using the asymptotics for small~$z$~\cite[(13.5.6)]{AS}, we find that
again after dropping the error term in~(\ref{calV}), $\grave{\psi}$ behaves
for small~$|\omega|$ as follows:
\[ \grave{\psi} \;=\; \omega^{-s-\sigma} r^{-\sigma}\;
\frac{(-4)^{-\frac{\sigma}{4}} \:\Gamma(2 \sigma+2)}{(2i)^s \Gamma(\sigma + 1-s)} \]
with~$\sigma$ as in~(\ref{sigmadef}).
This function satisfies~(\ref{philim}) and~(\ref{phi0asy}).

It remains to prove that the error term in~(\ref{calV}) does not destroy~(\ref{philim}, \ref{phi0asy}).
We first note that for~$r > K/|\omega|$ (with~$K$ as in
Theorem~\ref{thm41}), the WKB estimates of Section~\ref{sec4a} apply and show that~$\grave{\phi}$
is well-approximated by the above Whittaker functions. Let us next show that for some~$\delta>0$,
we can control the solution on the interval~$|\omega|^{-1+\delta} \leq r \leq K |\omega|^{-1}$. 
To this end, we introduce (similar to~\cite[Section~6]{FKSY}) the matrix
\[ A \;=\; \left( \!\!\begin{array}{cc} M_{\kappa, \mu}(2 i \omega r) & W_{\kappa, \mu}(2 i \omega r) \\
\partial_r M_{\kappa, \mu}(2 i \omega r) & \partial_r W_{\kappa, \mu}(2 i \omega r)
\end{array} \!\!\right) \]
and the function
\[ \Phi \;=\; A^{-1} \left( \!\!\begin{array}{c} \psi(r) \\ \psi'(r) \end{array}
\!\! \right)  . \]
Then~$\Phi$ satisfies the equation
\[ \Phi' \;=\; A^{-1} \left( \!\!\begin{array}{cc} 0 & 0 \\ {\mathcal{O}}(r^{-3}) & 0  \end{array}
\!\! \right) A \, \Phi \:. \]
Again using the asymptotic formulas~\cite[(13.1.32), (13.1.33), (13.5.1), (13.5.2)]{AS},
one finds that~$|\det A| \geq |\omega|/c$, and we obtain the inequality
\[ |\Phi|' \;\leq\;  \rho\, |\Phi| \quad {\mbox{where}} \quad 
\rho \;:=\; \frac{c}{|\omega|\, r^3}\: \|A\|^2 \:. \]
Applying Gronwall's inequalities
\begin{eqnarray*}
|\Phi(r_2)| &\leq& |\Phi(r_1)|\: \exp \left( \int_{r_0}^{r_1}
\rho \right) \\
|\Phi(r_2) - \Phi(r_1)|
&\leq& |\Phi(r_1)| \: \exp \left( \int_{r_0}^{r_1}\rho \right)\;
\int_{r_0}^{r_1}\rho\:,
\end{eqnarray*}
we can easily control~$\Phi$ provided that
the integral of~$\rho$ becomes arbitrarily small for small~$|\omega|$.
To see this, we first note that the Whittaker functions are bounded near~$z=0$
by~$c\: |z|^{\frac{1}{2}-\mu}$ and so~$\|A\|^2 \leq c^2\: |z|^{1-2 \mu}$
(see~\cite[(13.5.5) and (13.5.6)]{AS}). Thus
\[ \int_{|\omega|^{-1+\delta}}^{K |\omega|^{-1}} \rho(r)\, dr
\;\leq\; \frac{c^3}{|\omega|^{2 \mu}} \int_{|\omega|^{-1+\delta}}^{K |\omega|^{-1}}
\frac{dr}{r^{2+2 \mu}} \;\leq\;
\frac{c^3}{|\omega|^{2 \mu}} \;\frac{1}{1+2 \mu} \: |\omega|^{(1-\delta)(1+2 \mu)} \:, \]
and choosing~$\delta < (1+2 \mu)^{-1}$, the right side converges to zero as~$\omega \rightarrow 0$.

On the remaining interval~$r<|\omega|^{-1+\delta}$, we write the Schr\"odinger equation~(\ref{schV}) as
\begin{eqnarray*}
\lefteqn{ \left( -\partial_r^2 + \omega^2 - \frac{s^2+\lambda}{r^2} \right) \psi } \\
&=& \left[ {\mathcal{O}}(r^{-3}) + \left(- 2\, \frac{i s \omega + M \omega^2}{r} -
\frac{2 i M s \omega + 12 M^2 \, \omega^2}{r^2} \right) \Theta(|\omega|^{-1+\delta}-r) \right] \psi\:,
\end{eqnarray*}
where we used the Heaviside function to truncate the potential in the region which is of no
relevance here.
Treating the operator on the left as the free operator, its solutions are given by Hankel functions
(see~\cite{FKSY2, K}). A short calculation shows that the square bracket satisfies the condition
that~$\|r\, [\ldots]\|_{L^1}$ is bounded uniformly in~$|\omega|$. This is precisely
the condition which ensures the existence of the Jost solutions (see~\cite[Proof of Lemma~3.6]{FKSY2})
and again gives us control of the error terms.
Furthermore, one sees that in the region~$1 \ll r<|\omega|^{-1+\delta}$, the fundamental solutions are well-approximated by the Hankel functions, which in turn are a limiting case of our above Whittaker solutions.
This shows that the solution~$\grave{\psi}$ of the untruncated equation~(\ref{schV}, \ref{calV})
has a limit as~$\omega \rightarrow 0$, and that the asymptotics of the limit is the same
as that of the Whittaker solutions. This justifies dropping the error term in~(\ref{calV}).
\QED
Combining the last lemma with a convexity argument, we next show that
the Green's function has a limit at~$\omega=0$.
\begin{Lemma} \label{lemma82}
For the Green's function~$G(u,v)$ (\ref{Gdef}) of the Schr\"odinger
equation~(\ref{schroedinger}), the limit
\[ \lim_{C_\varepsilon(0) \ni \omega \rightarrow 0} G(u,v) \]
exists and is finite.
\end{Lemma}
{\Proof}
Using~(\ref{philim}, \ref{phi0asy}), the Green's function (\ref{Gdef}) has a limit at~$\omega=0$,
\[ \lim_{C_\varepsilon(0) \ni \omega \rightarrow 0} G(u,v) \;=\;
\frac{1}{w(\acute{\phi}, \grave{\phi}_0)}
\:\times\:
\left\{ \begin{array}{cl} \acute{\phi}(u)\, \grave{\phi}_0(v) & {\mbox{if~$v \geq u$}} \\
\grave{\phi}_0(u)\, \acute{\phi}(v) & {\mbox{if~$v < u$}}, \end{array} \right. \]
provided that the Wronskian on the right side does not vanish.
In order to show that this Wronskian is indeed non-zero, let us assume on the contrary that~$\acute{\phi}$ is a multiple of~$\grave{\phi}_0$.
Note that for~$\omega=0$, the potential~$V$ in~(\ref{Vdef})
is real and positive.
Hence we can repeat the convexity argument in the
proof of Lemma~\ref{lemmawcomplex} to get a contradiction,
where now we use that~$\phi_0$ tends to
zero at infinity according to~(\ref{phi0asy}) .
\QED

We next consider the case~$\omega_0 \neq 0$. We first show that~$\grave{\phi}$ has a well-defined limit
as~$\omega \rightarrow \omega_0$.
\begin{Lemma}
The following limit exists for any real~$\omega_0 \neq 0$ and every~$u \in \R$,
\[ \lim_{C_\varepsilon(\omega_0) \ni \omega \rightarrow \omega_0} \grave{\phi}(u) \;=\; \grave{\phi}_0(u) \:. \]
The limiting function~$\grave{\phi}_0$ is again a solution of the Schr\"odinger equation~(\ref{schroedinger})
with the asymptotics~(\ref{phiasy}).
\end{Lemma}
{\Proof} We cannot introduce~$\grave{\phi}_0$ directly via the iteration scheme~(\ref{Eind}) because
for real~$\omega$ the factor~$\exp(-2 \int_u^x \sqrt{V})$ is for large~$x$ increasing
polynomially like~$(x-u)^{2s}$. In order to bypass this problem, we introduce a convergence
generating factor; namely, we set for~$\omega = \omega_0$
\beq  \label{Eindmod}
\left. \begin{array}{rcl} E^{(0)} &\equiv& 1 \\[.5em]
E^{(l+1)}(u) &=& \displaystyle
\lim_{\delta \searrow 0} \int_u^\infty e^{-\delta x} \frac{W(x)}{2\, \sqrt{V(x)}}
\left\{ 1 - e^{-2 \int_u^x \sqrt{V}} \right\} E^{(l)}(x)\: dx \:,
\end{array} \right\}
\eeq
and define~$\grave{\phi}_0$ by
\beq \label{seriesmod}
\grave{\phi}_0(u) \;=\; \sum_{l=1}^\infty E^{(l)}(u)\: \grave{\alpha}(u)\:.
\eeq
Let us verify that this iteration scheme is well-defined and defines a solution of
the Schr\"odinger equation~(\ref{schroedinger}).
To this end, in~(\ref{Eindmod}) we substitute the identity
\[ e^{-2 \int_u^x \sqrt{V}} \;=\; \left(\frac{1}{-2 \sqrt{V}} \frac{d}{dx} \right)^p
e^{-2 \int_u^x \sqrt{V}} \:, \]
where, as in the proof of Theorem~\ref{thm41}, we choose~$p=0, 1, 3$ depending on whether~$s=\frac{1}{2}$, $1$ or~$2$,
respectively. After integrating by parts $p$ times, the resulting integrands are dominated by~$c/x^2$,
and thus we can take the limit~$\delta \searrow 0$ using Lebesgue's dominated convergence theorem.
The respective estimates~(\ref{se1}, \ref{se2}) and~(\ref{se3}, \ref{se4}) for~$s=1$ or~$s=2$
clearly remain valid for this modified iteration scheme, showing that the series~(\ref{seriesmod})
converges absolutely, uniformly for sufficiently large~$u$.

In order to compute the~$u$ derivative of~$E^{(l+1)}$, we integrate by parts,
take the limit~$\delta \searrow 0$, and can then compute the derivative.
After this, we can re-insert the convergence generating factor and re-integrate by parts.
This shows that in the formula for~$E^{(l+1)}$ we may interchange differentiation
with taking the limit~$\delta \searrow 0$. Exactly as above, one verifies that
the series~$\sum_{l=0}^\infty (E^{(l)})'$ converges again absolutely, uniformly for
sufficiently large~$u$. Hence~(\ref{seriesmod}) may be differentiated termwise, thereby
showing that~$\grave{\phi}_0$ is indeed a solution of~(\ref{schroedinger}).

Finally, to show continuity as~$C_\varepsilon(\omega_0) \ni \omega \rightarrow \omega_0$,
we first note that because of the continuous dependence of the solutions
of ODEs on both initial data and parameters on compact sets, it suffices to show 
continuity of~$\grave{\phi}(u)$ for~$u>u_1$ for any sufficiently large~$u_1$.
Again using the above integration-by-parts method, one sees that,
each of the~$E^{(l)}(u)$ is continuous as~$C_\varepsilon(\omega_0) \ni \omega \rightarrow \omega_0$.
Since for sufficiently large~$u_1$,
the series converges absolutely, uniformly in~$\omega \in C_\varepsilon(\omega_0)
\cup \{0\}$, we can take the termwise limit~$\omega \rightarrow \omega_0$.
\QED
From this lemma it will follow immediately that the Green's function converges,
\beq \label{Greenlimit2}
\lim_{C_\varepsilon(\omega_0) \ni \omega \rightarrow \omega_0} G(u,v) \;=\;
\frac{1}{w(\acute{\phi}, \grave{\phi}_0)} 
\:\times\:
\left\{ \begin{array}{cl} \acute{\phi}(u)\, \grave{\phi}_0(v) & {\mbox{if~$v \geq u$}} \\
\grave{\phi}_0(u)\, \acute{\phi}(v) & {\mbox{if~$v < u$}}, \end{array} \right.
\eeq
once we have shown that
the Wronskian~$w(\acute{\phi}, \grave{\phi}_0)$ is non-zero at~$\omega_0$.
This is done in the next lemma.
\begin{Lemma} \label{lemma83}
For any~$\omega_0 \neq 0$, the Wronskian~$w(\acute{\phi},
\grave{\phi}_0) \neq 0$.
\end{Lemma}
{\Proof} Assume that~$w(\acute{\phi}, \grave{\phi}_0)=0$.
We choose a function~$\eta \in C^\infty_0([-1,1])$.
For any~$\varepsilon < \omega_0/2$, we set
\[ \eta_\varepsilon(\omega) \;=\; \eta\!\left(\frac{\omega-\omega_0}{\varepsilon} \right) \]
and introduce the function
\beq \label{tphi}
R(t, u) \;=\; \frac{1}{r(u)}\, \int_\R d\omega\, e^{-i \omega t}\:\eta_\varepsilon(\omega)\: \phi_\omega(u) \:,
\eeq
where~$\phi_\omega(u)$ is the Jost solution~$\acute{\phi}$.
These Jost solutions have the following asymptotics,
\begin{eqnarray*}
\phi_\omega(u) &\sim& e^{\frac{s u}{4M}}\: e^{i \omega u} \spc\spc\spc\spc\quad\!\, {\mbox{as $u \rightarrow -\infty$}} \\
\phi_\omega(u) &\sim& c_1(\omega)\: u^s\: e^{-i \omega u} \:+\:
c_2(\omega)\: u^{-s}\: e^{i \omega u} \qquad  {\mbox{as $u \rightarrow \infty$}}\:,
\end{eqnarray*}
where~$c_1$ and~$c_2$ depend smoothly on~$\omega$ and~$c_2(\omega_0)=0$. Thus for any~$\delta>0$
we can choose~$\varepsilon$ such that
\[ |c_2(\omega)| \;\leq\; \delta \qquad \forall \omega \in B_\varepsilon(\omega_0)\:. \]

Differentiating~(\ref{tphi}) and using that~$\phi_\omega$ are solutions of the Schr\"odinger
equation~(\ref{schroedinger}), one easily verifies that~$R(t,u)$ is a solution of the
Teukolsky equation~(\ref{teukolskysep}) (with the angular dependence separated out).
Near the event horizon, $R(t,u)$ has the following asymptotics,
\beq \label{asyevent}
R(t,u) \;\sim\; \frac{e^{\frac{s u}{4M}}}{r} \int_\R d\omega\, e^{-i \omega t}\:\eta_\varepsilon(\omega)\: e^{i \omega u}
\;=\; \frac{e^{\frac{s u}{4M}}}{r}\: \hat{\eta}_\varepsilon(t-u)\:,
\eeq
where~$\hat{\eta}_\varepsilon$ denotes the Fourier transform of~$\eta_\varepsilon$. Similarly, near infinity,
\beq \label{asyinfinity}
R(t,u) \;\sim\; \frac{u^s}{r} (\widehat{c_1 \eta_\varepsilon})(t+u) + \frac{u^{-s}}{r} (\widehat{c_2
\eta_\varepsilon})(t-u)\:.
\eeq
Being the Fourier transform of a smooth function supported in~$B_\varepsilon(\omega_0)$,
the functions~$\hat{\eta}$, $\widehat{c_1 \eta_\varepsilon}$ and $\widehat{c_2 \eta_\varepsilon}$ all have rapid
decay on the scale~$\varepsilon^{-1}$, i.e.
\[ \sup_{x} |x|^n
\left( |\hat{\eta}_\varepsilon| + |\widehat{c_1 \eta_\varepsilon}| + |\widehat{c_2 \eta_\varepsilon}|
\right)(x) \;\leq\; \frac{c_n}{\varepsilon^n}\:. \]
Furthermore, the function~$(\widehat{c_2 \eta_\varepsilon})$ is pointwise small,
\[ |(\widehat{c_2 \eta_\varepsilon})(t-u)| \;\leq\; \left| \int_\R d\omega\, e^{-i \omega (t-u)}\:c_2(\omega)\:
\eta_\varepsilon(\omega) \right| \;\leq\; \delta\: \|\eta_\varepsilon\|_{L^1}\:; \]
similarly, all its derivatives are pointwise small.
The formulas~(\ref{asyevent}) and~(\ref{asyinfinity}) are valid near~$u=-\infty$ and~$u=\infty$, respectively.
Since~$\omega$ is in a compact set, the error terms in the asymptotics are bounded uniformly in time.

The asymptotics~(\ref{asyevent}, \ref{asyinfinity}) contradict the conservation of physical energy.
Namely, for large negative times, (\ref{asyevent}) describes a wave of positive energy coming from the
event horizon. However, for large positive times, the contribution of~(\ref{asyevent}) as
well as the first summand in~(\ref{asyinfinity}) decay rapidly in time, whereas
the energy of the second summand in~(\ref{asyinfinity}),
which describes a wave moving to infinity, can be made arbitrarily small by choosing~$\delta$ small.
\QED
We remark that if the above energy argument is made more quantitative, it even yields that
\[ \left| \frac{c_1}{c_2}\right| \;\leq\; 1\:. \]
This is in complete agreement with the numerical result~$Z \leq 1$ in the case~$a/M=0$
obtained in~\cite[p.~658ff]{PT}, keeping in mind that, after a time reflection, the
quantities $Z_{\mbox{\scriptsize{in}}}$ and~$Z_{\mbox{\scriptsize{out}}}$ as introduced in~\cite{PT}
are multiples of our coefficients~$c_2$ and~$c_1$, respectively.

Using Lemmas~\ref{lemma82} and~\ref{lemma83}, for every~$\omega_0 \in \R$,
we can introduce the function~${\mathfrak{R}}_{\omega_0}$ as the limit of the integral kernel
of the resolvent from the lower half plane; namely,
\[ {\mathfrak{R}}_{\omega_0}(u,v) \;:=\; \lim_{C_\varepsilon(\omega_0) \ni \omega \rightarrow \omega_0}
R_\omega(u,v)\:. \]
In Theorem~\ref{thmintrep1} we can take the limit~$R \rightarrow \infty$
and deform the lower contour~$\mathfrak{C}_1$ onto the real axis, the upper
contour~$\mathfrak{C}_2$ onto the line~${\mbox{Im}} \,\omega = \frac{s}{2M}$,
to obtain the following integral representation, valid for all~$t \in \R$.
\begin{Thm} \label{thmintrep2} For any spin $s \in \{\frac{1}{2}, 1, 2\}$,
the solution of the Cauchy problem for the separated Teukolsky equation~(\ref{teq2})
with initial data~$\Psi_0 = (\phi, \partial_t \phi)|_{t=0} \in C^\infty_0(\R)^2$
can be written as
\begin{eqnarray*}
\Psi(t,u) &=& -\frac{1}{2 \pi i} \int_{\sR} e^{-i \omega t}
\left( ({\mathfrak{R}}_\omega \Psi_0)(u) + \frac{\Psi_0(u)}{\omega+i} \right) d\omega \\
&&+\frac{1}{2 \pi i} \int_{\sR + \frac{i s}{2M}}  e^{-i \omega t} \left(
(R_\omega \Psi_0)(u) + \frac{\Psi_0(u)}{\omega+i} \right) d\omega\:,
\end{eqnarray*}
where both integrals are $L^1$-convergent.
\end{Thm}
{\Proof} As shown in the proof of Theorem~\ref{thmintrep1}, inserting the term~$\Psi_0(u)/(\omega+i)$
into the integrand in~(\ref{intrep}) does not change the value of the limit.
According to~(\ref{Rident}), the resulting integrand is bounded near infinity by~$C/|\omega|^2$,
and hence we can take the limit~$R \rightarrow \infty$ in the Lebesgue sense.
\QED

\section{Proof of Decay} \label{secdecay}
\setcounter{equation}{0}

We now prove our main theorem. \\[1em]
{\em{Proof of Theorem~\ref{thm1}. }}
As discussed in the introduction, the conservation of energy implies that the
solution~$\Phi$ of the Cauchy problem~(\ref{teq}, \ref{id}) is bounded in~$L^2_{\mbox{\scriptsize{loc}}}$,
uniformly in time.
Differentiating the Teukolsky equation with respect to~$t$, one sees that the derivatives~$\partial^n_t \Phi$
are also solutions and are thus also bounded in~$L^2_{\mbox{\scriptsize{loc}}}$.
Since the spatial part of the Teukolsky equation~(\ref{teq}) is uniformly elliptic away from the event horizon, we conclude that all spatial derivatives of~$\Phi$ are bounded in~$L^2_{\mbox{\scriptsize{loc}}}$.
Using the Sobolev embedding~$H^{2,2}_{\mbox{\scriptsize{loc}}} \hookrightarrow L^\infty_{\mbox{\scriptsize{loc}}}$,
we conclude that~$\Phi$
can be bounded in~$L^\infty_{\mbox{\scriptsize{loc}}}$, uniformly in time, by
a Sobolev norm of the initial data, i.e. for any compact set~$K \subset (r_1, \infty) \times S^2$,
\beq \label{sobin}
\sup_K |\Phi(t)| \;\leq\;
c \; \|(\Phi_0, \Phi_1)\|_{H^{2,2}}\:,
\eeq
where~$c$ depends only on~$K$ and the support of the initial data.

Decomposing the initial data into spin-weighted spherical harmonics~\cite{spinweight},
\[ (\Phi_0, \Phi_1)(r, \vartheta, \varphi) \;=\; \sum_{l=s}^\infty
\sum_{m=-l}^l 
\:_sY_{lm}(\vartheta, \varphi)\:(\Phi_0^{l,m}, \Phi^{l,m}_1)(r, \vartheta, \varphi)\:, \]
the Sobolev norm decomposes into a sum over the angular momentum modes,
\[ \|(\Phi_0, \Phi_1)\|^2_{H^{2,2}} \;=\; \sum_{l,m} \left\|
\:_sY_{lm} \:(\Phi_0^{l,m}, \Phi^{l,m}_1)
\right\|^2_{H^{2,2}}\:. \]
Since the series converges absolutely, for any~$\varepsilon>0$ there is an integer~$l_0$ such that
\[ \sum_{l > l_0} \;\sum_{m=-l}^l \left\|\:_sY_{lm}\:(\Phi_0^{l,m},
\Phi^{l,m}_1) \right\|^2_{H^{2,2}} \;\leq\; \varepsilon\:. \]
Hence in view of~(\ref{sobin}), the contribution of the large angular momentum modes to~$|\Phi|$ can be
made pointwise small, uniformly in time.

For the remaining finite number of angular momentum modes, we use the integral representation
of Theorem~\ref{thmintrep2}. The integral on the real line tends to zero 
as~$t \rightarrow -\infty$ by virtue of the Riemann-Lebesgue lemma.
The integral on the line~${\mbox{Im}} \,\omega=\frac{s}{2M}$ 
can be bounded by a constant times~$\exp(\frac{st}{2M})$ and thus tends to zero exponentially fast as~$t \rightarrow -\infty$.
\QED

\section{General Remarks} \label{sec6}
\setcounter{equation}{0}
We first discuss the case~$s=\frac{1}{2}$ of the massless Dirac equation.
At first sight, it might seem paradoxical that in this case,
the solution of the Cauchy problem has two different integral representations,
one being the representation obtained in~\cite{FKSY03} where the $\omega$-integral runs
over the real axis, the other being that given in Theorem~\ref{thmintrep2}, where $\omega$
is integrated over two lines in the complex plane.
This, however, is no contradiction, as one can understand as follows.
In~\cite{FKSY03} the massless Dirac equation is considered as a first-order system.
The Teukolsky equation, on the other hand, is a second order equation
for a single component of the Dirac system. It is obtained from the Dirac equation
by multiplying with a particular first-order operator.
The transformation from the Dirac equation to the Teukolsky equation completely changes the spectrum of
the involved operators. Whereas the Hamiltonian of the Dirac system
is self-adjoint and thus has a purely real spectrum, the Hamiltonian of the Teukolsky
system has non-real essential spectrum (see Proposition~\ref{prpess}).
The differences in the integral representations reflect these differences in the spectra
of the corresponding Hamiltonians.

It is important to note that the differences in the spectral representations do not
imply different long-time dynamics. To explain this better, let us consider
the integral representation of Theorem~\ref{thmintrep2} in the limit~$t \rightarrow +\infty$.
In this limit the exponential factor~$e^{-i \omega t}$ in the second integral
grows exponentially in time, suggesting that~$\Psi(t,u)$ should also increase in time.
However, this reasoning is not valid because, as explained in the introduction, by considering
the time-reflected Teukolsky equation for~$-s$ and using the Teukolsky-Starobinsky identities,
we conclude that~$\Psi(t,u)$ indeed also decays as~$t \rightarrow +\infty$.
Another way of seeing why the naive reasoning is not valid is to consider the asymptotics
near $u=+\infty$. Then the the fundamental solutions~$\grave{\phi}(u)$
appearing in the resolvent of the second integral, decay exponentially as~$u \rightarrow
\infty$. This leads to an exponential damping of an outgoing wave moving towards infinity,
which just compensates the exponential increase of the factor~$e^{-i \omega t}$. This
argument illustrates how Theorem~\ref{thmintrep2} describes the correct dynamics
in the asymptotic limit of wave packets near spatial infinity.

We end the paper by discussing to which extent our arguments carry over to the
Kerr metric. The first complication in Kerr is that the separation constant~$\lambda$
depends now on~$\omega$, and thus the sum of all angular momentum modes
must be carried along at each step, and this would require additional estimates
to control the infinite sum. Apart from this additional complication,
our arguments in Sections~\ref{sec2}--\ref{seccom} continue to hold.
In Section~\ref{seccontour} the considerations before Lemma~\ref{lemma83}
could also be extended, provided that the sum over the angular momentum modes can
be controlled. However, the energy argument of Lemma~\ref{lemma83} no longer
works in the Kerr geometry due to the presence of the ergosphere, where the
physical energy density need not be positive.
The numerics carried out by Press and Teukolsky~\cite{PT} indicate that even in the
Kerr geometry, the Wronskian~$w(\acute{\phi}, \grave{\phi}_0)$ has no
zeros on the real line.
A possible strategy to make this rigorous would be to replace 
our above energy argument by a
causality argument in the spirit of~\cite[Section~7]{FKSY2}.
However, this would make it necessary to analytically extend the resolvent
across the real line. In principle, this could be achieved by
estimating the higher~$\omega$-derivatives of~$\grave{\phi}$ similar
to~\cite[Lemma~3.4]{FKSY2}. However, this approach seems to be technically very
demanding. \\[.5em]

\noindent
{\em{Acknowledgments:}} We would like to thank Bernard Whiting for
his very careful reading of this paper and for several valuable comments.
We are grateful to the Alexander-von-Humboldt foundation for its generous support. We would like to thank Thierry Daud{\'e},
Niky Kamran and Shing-Tung Yau for helpful disussions.

\begin{tabular}{lcl}
\\
Felix Finster & $\;\;\;\;$ & Joel Smoller \\
NWF I -- Mathematik && Mathematics Department \\
Universit{{\"a}}t Regensburg && The University of Michigan \\
93040 Regensburg, Germany && Ann Arbor, MI 48109, USA \\
{\tt{Felix.Finster@mathematik.uni-r.de}} && {\tt{smoller@umich.edu}}
\end{tabular}

\end{document}